\documentclass[12pt]{article}
\renewcommand{\baselinestretch}{1.6}
\begin{document}
\newcommand{\be}{\begin{equation}}
\newcommand{\ee}{\end{equation}}
\newcommand{\al}{\alpha}
\newcommand{\bt}{\beta}
\newcommand{\g}{\gamma}
\newcommand{\G}{\Gamma}
\newcommand{\ve}{\varepsilon}
\title{
\vspace*{-50mm}
\begin{flushright}
{\large\bf MZ-TH/98-61}\\[-5mm]
{\large December 1998}
\end{flushright}
{ \Large \bf Analytical results for $e^+e^-\rightarrow t\bar t$ and
$\g\g\rightarrow t\bar t$
observables near the threshold up to the next-to-next-to-leading order
of NRQCD.
}}
\author{
  {\bf A.A.Penin}\thanks{On leave from Institute for Nuclear Research,
  Moscow, Russia}\\
  {\small {\em Institut f{\"u}r Theoretische Teilchenphysik
  Universit{\"a}t Karlsruhe}}\\[-3mm]
  {\small {\em D-76128 Karlsruhe, Germany}}\\
 {\bf A.A.Pivovarov}{$^*$}\\
  {\small {\em
Institut f\"ur Physik, Johannes-Gutenberg-Universit\"at}}\\[-3mm]
  {\small {\em  D-55099 Mainz, Germany}}}

\date{}

\maketitle

\begin{abstract}
We present the  analytical description of
 top-antitop pair production near the threshold in
$e^+e^-$ annihilation and $\g\g$ collisions.
A set of basic observables considered
includes the total cross sections, forward-backward
asymmetry and top quark polarization.
The threshold effects relevant for the basic observables
are described by three universal
functions related to $S$ wave production, $P$ wave production and
$S-P$ interference. These functions are computed analytically
up to the next-to-next-to-leading order of NRQCD.
The total $e^+e^-\rightarrow t\bar t$ cross section
near the threshold
is obtained in the  next-to-next-to-leading order
in the closed form including the contribution
due to the axial coupling
of top quark and mediated by the $Z$-boson. The effects
of the running of the strong
coupling constant and of the finite top quark width
are taken into account analytically
for the $P$ wave
production and $S-P$ wave interference.
\\[2mm]
\end{abstract}
\thispagestyle{empty}

\newpage
\section{Introduction.}
Being heavy the top quark undergoes fast weak decays.
The relatively large
width $\G_t$ of the top quark is mainly saturated by the decay channel
$t\rightarrow Wb$  and keeps
the effective energy of top-antitop system in the complex plane
far enough from the cut along the positive semiaxis.
Thus it serves as a sufficient infrared cutoff for long distance
effects avoiding the problem of strong coupling.
This allows one to bypass possible nonperturbative
regions and is the key observation
for
the theoretical study of the top-antitop
pair production near the two-particle threshold
\cite{FK}.
Because the relevant scale
$\sqrt{\G_tm_t}$, with $m_t$ being the top quark mass, is much larger
than $\Lambda_{\rm QCD}$, the QCD perturbation theory expansion
is applicable for the theoretical
description of physical phenomena near the
top quark threshold if singular  Coulomb effects are  properly taken into
account \cite{FK,ttgg,ttee}.
This feature turns the processes involving the top quarks into a unique
laboratory for perturbative investigation of threshold effects.
Experimental study of the top-antitop
pair threshold production is planned to
be performed at the Next Linear Collider
both in high energy $e^+e^-$ annihilation and
$\gamma\gamma$ collision \cite{exp}.
High quality experimental data that can be obtained in such experiments
along with a very accurate theoretical description of them
make the
processes of top-antitop
pair threshold production a promising place
for investigating quark-gluon
interactions.
This investigation concerns both general features
of interaction and precise quantitative properties such as
the determination of numerical values of
the strong coupling constant $\al_s$,
the top quark mass, and the top quark width.
Though the main features in both
$e^+e^-$ and
$\gamma\gamma$
processes of top
quark pair threshold production  are rather similar
the strong interaction corrections and relativistic
corrections are different for them.
Therefore a simultaneous analysis of
these two processes extends possibilities of studying
fine details of the top quark threshold dynamics.
Besides the total cross sections which are mainly saturated by
the $S$ wave final state of the top quark-antiquark pair,
there is a set of observables sensitive to the $P$ wave
component. For example, the $S$ and $P$
partial waves of the final state top quark-antiquark pair
produced in $\gamma\gamma$ collisions can be separated
by choosing the same or opposite helicities of the
colliding photons \cite{ttgg}. This gives an opportunity of direct
measurement of the $P$ wave amplitude which is strongly
suppressed in the threshold
region in comparison with the $S$ wave one. On the other hand,
the forward-backward
asymmetry of the  quark-antiquark pair
production in $e^+e^-$ annihilation \cite{Sum,Har} and
top quark polarization \cite{Har,FKK} are
determined by the $S-P$ wave interference
in both processes.
This provides us with two additional
independent probes
of  the  top quark interactions.

The finite order perturbation theory of QCD
breaks down in the threshold region of particle production
due to the presence of singular $(\al_s/\bt)^n$
Coulomb terms. Here $\bt$ is a velocity of the heavy quark.
However the resummation of these Coulomb contributions
which are most important quantitatively in the threshold region
is
possible and can be
systematically done in the framework of nonrelativistic
QCD (NRQCD) \cite{CasLep} (for the recent development
of the NRQCD effective theory approach see  
\cite{Man,LukMan,GriRot,LukSav,PinSot,BenSmi,Lab,Gr}). 
Note that the characteristic scale
of the Coulomb effects for the top quark production $\al_s m_t$ is
comparable numerically with the cutoff scale for infrared
effects
$\sqrt{\G_t m_t}$ so
the Coulomb effects are not suppressed by the top quark width.
The determination of higher order corrections
in the QCD coupling constant
and of relativistic corrections in the case
when Coulomb effects have to be taken into account beyond the finite
order perturbation theory
requires the perturbative expansion for the complete
correlator to be performed
near the Coulomb approximation rather than near Green functions
of the free theory
which is the standard pattern of
perturbation theory for the infrared safe high energy processes.

Recently has been made a rather
essential
progress in the theoretical description of
heavy quark-antiquark threshold dynamics
within NRQCD. The evaluation of next-to-leading order (NLO)
and next-to-next-to-leading order (NNLO)
corrections to the heavy quark threshold production in $e^+e^-$
annihilation has been done
both within the analytical approach
\cite{H1,KPP,PP,MY,PP1,BSS,BenSin,H2,KniPen}
and numerically \cite{Hoang,Mel,KuhTeu,Nag,HoaTeu}
while
the NLO corrections to the heavy quark threshold production in
$\gamma\gamma$ collision where computed analytically \cite{PP2}.
The analysis
of NNLO corrections in the last case is still absent.
However, this analysis is necessary for the accurate
quantitative study of the process since the NNLO contribution
is found to be relatively large in the case of the top quark production
in $e^+e^-$ annihilation \cite{Hoang,Mel} and one can expect
that some large corrections emerge also in the case of the top quark
threshold production in $\gamma\gamma$ collision.
Moreover, a semianalytical analysis
of the high order corrections to
the top quark threshold production cross section in $e^+e^-$ annihilation
has been performed so far \cite{Hoang,Mel,Nag,HoaTeu}  while
the essential part of corrections has been accounted for
numerically \cite{ttee}. Therefore the complete
analytical description of the
process is also desirable\footnote{When this work was in its final stage 
a letter \cite{BSS} appeared where the photon mediated  top quark 
production in $e^+e^-$ annihilation was analyzed analytically.}.
Furthermore,
the forward-backward asymmetry and top
quark polarization has been analyzed
in NLO only numerically \cite{Har}
as well as the axial contribution in
the $e^+e^-\rightarrow t\bar t$ process \cite{KuhTeu}.
In this case the numerical study is more involved because of
necessity to construct the Green function
for the $P$ wave which leads to
the more singular differential equations
in comparison with the $S$ wave.
The case of
$P$ wave production in $\gamma\gamma$ collision \cite{ttgg,PP2}
clearly demonstrates that
the  numerical analysis  \cite{KuhTeu} with an explicit cutoff of
the hard  momentum contribution is not sufficient for
an accurate account for the finite top quark width for
these quantities because the relativistic effects
are not taken into account properly.
   
In the present paper we give a simultaneous analysis
of several observables relevant to
$e^+e^-\rightarrow t\bar t$ annihilation and $\g\g\rightarrow t\bar t$
collisions near the top quark production
threshold in high orders of NRQCD.
The total cross sections are computed in NNLO
of NRQCD which includes
$\al_s^2$, $\al_s\bt$ and $\bt^2$ corrections
in the coupling constant $\al_s$
and the heavy quark velocity $\bt$
to the nonrelativistic Coulomb approximation.
Explicit analytical expressions for the soft part of
corrections are obtained.
The threshold cross section of the $t \bar t$ production in
$e^+e^-$ annihilation
is obtained in the closed form including the contribution
due to the top quark axial coupling.
The hard part of the correction to
the $\g\g\rightarrow t\bar t$ threshold cross section is found
with the logarithmic accuracy. The forward-backward
asymmetry of the top quark-antiquark pair
production in  $e^+e^-$ annihilation  and top quark polarization
in both processes of $e^+e^-$ annihilation
and $\g\g$ collisions
are computed up to NLO.

The paper is organized as follows.
In the next Section the nonrelativistic approximation for the
basic observables of top  quark-antiquark pair
production near the threshold is formulated.
In Section 3 the threshold effects are described by three universal
functions related to the $S$, $P$ wave production and
$S-P$ wave interference which have been computed analytically
within NRQCD.
In Section 4 we present our numerical analysis
and the discussion of the obtained results. The last Section is
devoted to our conclusions.  Some explicit analytical formulae are given
in Appendix.

\section{The nonrelativistic approximation near
the production threshold.}
In this section we describe the set of observable which
will be analyzed: the total cross sections, the forward-backward asymmetry,
the polarization of top quark. 
We formulate the nonrelativistic approximation for
these observables setting the stage for the complete NRQCD analysis. 
In the last subsection we dwell upon the
peculiarities of introducing the finite width of the top quark.

\subsection{The effective theory description of the 
heavy quark threshold dynamics.}

Near the threshold the heavy quarks are nonrelativistic so that one may 
consider both the strong coupling constant and 
heavy quark velocity   as  small parameters.
The threshold expansion of the  QCD loop integrals 
has been developed in  \cite{BenSmi}. 
However, to take into account the singular threshold effects 
properly one  has to go beyond the  finite order QCD 
perturbation theory. For this purpose the  expansion in $\bt$ should 
be performed directly in the QCD Lagrangian within the effective 
field theory framework.
The first step to construct the effective theory is to identify
all the scales present in the problem. 
The  threshold dynamics is characterized by  four different scales 
\cite{BenSmi}:\\
(i) the hard scale (energy and momentum scale like $m_q$);\\
(ii) the soft scale (energy and momentum scale like $\bt m_q$);\\
(iii) the potential scale (the energy scales like $\bt^2 m_q$, while 
the momentum scales like $\bt m_q$); \\
(iv) the ultrasoft scale (both energy and momentum scale like $\bt^2 m_q$).
The ultrasoft scale is only relevant for gluons.\\
By integrating out the hard scale of QCD one arrives at the effective theory
of NRQCD \cite{CasLep}.
Because the NRQCD Lagrangian does not contain explicitly  the heavy
quark velocity  
the power counting rules are necessary  to construct the regular 
expansion in this parameter. 
The list of the  power counting rules for dimensionally
regularized NRQCD and their relation to the threshold 
expansion \cite{BenSmi} can be found in \cite{Gr}
Integrating out the soft modes and the potential gluons of NRQCD one 
obtains the effective theory of potential NRQCD \cite{PinSot} which contains
potential quarks and ultrasoft gluons as active particles 
and is relevant for the analysis of the threshold effects.
In potential NRQCD the dynamics of the quarks is governed by the effective 
nonrelativistic Schr{\"o}dinger equation and by their interaction 
with the ultrasoft gluons. To obtain a regular perturbative expansion 
in $\bt$ this interaction should be expanded in multipoles.   
Note that in the process of scale separation some spurious infrared and 
ultraviolet divergences may appear at intermediate steps of calculation
which cancel each other in the final results for physical observables.    
The dimensional regularization has been recognized as a powerful
tool to deal with these divergences 
\cite{Man,LukSav,PinSot,BenSmi,Lab,Gr,c1,c2,PinSot1,CMY}. 

If the ultrasoft effects are neglected 
the propagation of a  quark-antiquark pair  
in the color singlet state is described in the potential NRQCD
by the Green function $G({\bf x},{\bf y},E)$ of the  Schr{\"o}dinger equation
\be
\left({\cal H}-E\right)G({\bf x},{\bf y},E)=\delta({\bf x}-{\bf y})
\label{Schr}
\ee
where ${\cal H}$ is the effective nonrelativistic Hamiltonian.
Near the threshold  the singular $(\al_s/\bt)^n$
Coulomb terms should be summed up in all orders in $\al_s$. 
Thus, in threshold region one has to develop the expansion   
in $\bt$ and $\al_s$ around some 
solution which incorporates properly the threshold effects, for example,  
around the nonrelativistic Coulomb solution.
In this case the leading order approximation for the
nonrelativistic Green function 
is obtained with the Coulomb Hamiltonian
\[
{\cal H}_C=-{{\bf \Delta}_{\bf x}\over m_t}+V_C(x)
\]
where
${\bf \Delta_x} = \partial_{\bf x}^2$
is the kinetic energy operator 
and  $V_C(x)=-C_F\al_s/x$ is the Coulomb potential,
$x=|{\bf x}|$.
The  harder scales contributions  
are represented by the  
higher-dimension operators in ${\cal H}$   
and by the Wilson coefficients
of the operators of the nonrelativistic Hamiltonian  
leading to an expansion in $\bt$ and $\al_s$.
On the other hand 
the radiation/absorption of the ultrasoft gluons
by the interacting quark-antiquark pair, the effect of retardation, 
does not contribute in NLO and NNLO 
(the leading ultrasoft effects in heavy quarkonium 
have been recently computed in ref.~\cite{KniPen}). 
Thus, the nonrelativistic Green function of 
eq.~(\ref{Schr}) is the basic object in  NRQCD analysis 
of the threshold effects up to NNLO.
In Sections~2.2-2.4 we relate the  observables of  
$e^+e^-\rightarrow t\bar t$ annihilation and $\g\g\rightarrow t\bar t$
collisions  in the threshold region to this  Green function. 

\subsection{Cross sections.} We study the normalized cross
sections of the top quark-antiquark pair production in $e^+e^-$
annihilation
\[
R^e(s)={\sigma(e^+e^-\rightarrow t\bar t)\over
\sigma(e^+e^-\rightarrow\mu^+\mu^-)},
\]
and in $\g\g$ collisions
\[
R^\g(s)={\sigma(\gamma\gamma\rightarrow t\bar t)\over
\sigma(e^+e^-\rightarrow\mu^+\mu^-)}
\]
where the lepton cross section
\[
\sigma(e^+e^-\rightarrow\mu^+\mu^-)={4\pi\al_{QED}^2\over 3s}
\]
is the standard normalization factor
with $\al_{QED}$ being the fine structure constant.
Here $\sqrt{s}$ is the total energy of the colliding 
particles (electrons or photons) in the
center of mass frame.
For unpolarized initial states
the following decomposition
of the total cross sections
is useful
\be
R^{e}(s)={D_{V}\over q_t^2}R^{v}(s)+D_{A}R^{a}(s),
\label{Ree}
\ee
\be
R^{\g}(s)={R^{++}(s)+R^{+-}(s)\over 2}
\label{Rgg}
\ee
where $R^v$ ($R^a$) corresponds to the
top quark vector (axial) coupling
in $e^+e^-$ annihilation
while $R^{++}$ ($R^{+-}$) corresponds
to the colliding photons of the same (opposite) helicity
in $\g\g$ collisions.
$D_{V,A}$ are the standard combinations of electroweak coupling
constants (see below), $q_t$ is the top quark electric charge.
The cross section for the polarized electron/positron
initial states is discussed, e.g. in ref.~\cite{Kuh}.

Near the threshold the cross sections are determined by the imaginary part
of the correlators of the nonrelativistic vector/axial  
quark currents which can be related to the nonrelativistic
Green function and its derivatives at the origin.
In NNLO the (potential) NRQCD provides one with the 
following representation of the cross sections 
\be
R^{v}(s) ={6\pi q_t^2N_c\over  m_t^2}\left(
C^{v}(\al_s)+B^{v}{k^2\over m_t^2}\right){\rm Im}G(0,0,k),
\label{Rv}
\ee
\be
R^{a}(s)={4\pi N_c\over m^4}C^a(\al_s)
\partial^2_{\bf xy}
{\rm Im}G({\bf x},{\bf y},k)
\label{Ra}
\ee
\be
R^{++}(s)
={24\pi q_t^4N_c \over m_t^2}
\left(\left(C^{++}(\al_s)+B^{++}{k^2\over m_t^2}\right)
{\rm Im}G(0,0,k)+\partial^2_{\bf x\bf y}
{\rm Im}G({\bf x},{\bf y},k) \right)
\label{Rpp}
\ee
\be
R^{+-}(s)={32\pi q_t^4N_c\over m^4}C^{+-}(\al_s)
\partial^2_{\bf x\bf y}
{\rm Im}G({\bf x},{\bf y},k)
\label{Rpm}
\ee
where $k^2=-m_tE$,
$E=\sqrt{s}-2m_t$ is the energy of a quark pair
counted from the threshold $2m_t$.
A symbolic notation  $\partial^2_{\bf x\bf y}$ 
is used for the operator  
\[
\partial^2_{\bf x\bf y}f({\bf x},{\bf y})\equiv\sum_{i=1}^3\left.
\partial_{x_i}\left(\partial_{y_i}
f({\bf x},{\bf y})|_{y=0}\right)\right|_{x=0}
\]
that singles out the $P$ partial wave of the Green function.
The standard electroweak factors
read
\[
D_V=q_e^2q_t^2+2q_eq_tv_ev_td+(v_e^2+a_e^2)v_t^2d^2,
\]
\[
D_A=(v_e^2+a_e^2)a_t^2d^2,
\]
\[
\begin{array}{lll}
q_e=-1,         & v_e=-1+4\sin^2\theta_W ,& a_e=-1,\\
q_t={2\over 3}, & v_t=1-8\sin^2\theta_W  ,& a_t=1,
\end{array}
\]
\[
d={1\over 16\sin^2\theta_W\cos^2\theta_W}{s\over s-M_Z^2}.
\]
The coefficients $C^i(\al_s)$ and $B^i$ are the parameters of 
NRQCD which  are responsible for matching
the effective and full theory cross sections in the limit of weak
coupling in  NNLO. 
The coefficients
\[
C^{i}(\al_s)=1+c^{i}_1C_F{\al_s\over
\pi}+c^{i}_2C_F \left({\al_s\over \pi}\right)^2+\ldots
\]
account for
the hard QCD corrections and are determined by the corresponding 
amplitudes with on-shell heavy quarks at rest. The numerical values of these
hard coefficients in the NLO approximation have been known
since long ago \cite{Kar,RRY,HarBr,BarKuh}. They are explicitly
\[
c^v_1=-4,\qquad c^a_1=-2, 
\] 
\[ c^{++}_1={\displaystyle {\pi^2\over
4}-5},\qquad c^{+-}_1=-4\, .
\]
The coefficient $C^v$ has recently been computed
in NNLO in different schemes \cite{BSS,Hoang,Mel}.
Starting from NNLO the  hard  coefficients
acquire the anomalous dimensions and the calculation of the
NNLO correction requires an accurate separation of hard and soft
contributions. At the same time these  coefficients  
do not depend on  the normalization point of the strong coupling 
constant  in NNLO
and  one can use the different normalization points of 
$\al_s$ entering in the coefficients $C^i$ (the ``hard'' 
scale $\mu_h$) and the nonrelativistic Green function
(the ``soft'' scale $\mu_s$),  see Section~3.1. 

The coefficients $B^i$ in eqs.~(\ref{Rv}) and (\ref{Rpp})
describe the pure relativistic corrections to the cross section
which appear when the cross section is evaluated
in terms of the correlator of nonrelativistic quark
currents. Because the corresponding correction first
appears in the order
$O(\bt^2)$ the coefficients $B^i$
can be taken in the leading order in
$\al_s$. The coefficient $B^v$ is related to the
nonrelativistic expansion of the vector current and is known \cite{CasLep}
to be equal to
\[
B^v={4\over 3}.
\]
The calculation of the coefficient
$B^{++}$ necessary for the consistent description of the
$\g\g$ cross section
within
NRQCD in NNLO is more involved because the amplitude of
$\g\g\rightarrow t\bar t$ transition is determined by the
nonrelativistic expansion of a $T$ product of two vector currents
\cite{PP2,Piv}. This coefficient, however, can be found by direct
comparison with the relativistic expression for the cross section
expanded in the velocity of the heavy quark (see
Section~3.1).

For the noninteracting quarks (the Born approximation)
one obtains the following results for the cross sections 
$(\beta=\sqrt{1-4m_t^2/s})$
\[
R^v(\beta)={3\over 2}q_t^2N_c\theta(\beta^2)(\beta+O(\beta^3)),\qquad
R^a(\beta)=N_c\theta(\beta^2)(\beta^3+O(\beta^5)),
\]
\[
R^{++}(\beta)=6q_t^4N_c\theta(\beta^2)(\beta+O(\beta^3)),\qquad
R^{+-}(\beta)=8q_t^4N_c\theta(\beta^2)(\beta^3+O(\beta^5)).
\]
Note that the cross sections $R^v$ and $R^{++}$
are saturated with the $S$ wave contribution and are 
proportional to the Green function at the origin while $R^a$ and $R^{+-}$
parts are saturated with the $P$ wave contribution and 
are proportional to the derivative of the Green function at
the origin.
As a consequence they
are suppressed in comparison with $R^v$ and $R^{++}$ by
factor $\beta^2$. In the present paper we study the corrections
to the total
cross sections $R^e$ and $R^\g$ up to the NNLO of NRQCD.
Thus $R^a$ is a NNLO contribution to the total cross
section $R^e$ and only the leading contribution to $R^a$ is
important. On the contrary, the $R^{+-}$ part can be separated from $R^\g$
by fixing the opposite helicities of the colliding photons.
This makes possible the direct study of
the $P$ wave production and, therefore,
the evaluation of the
corrections to $R^{+-}$ cross section is of practical interest.

Concluding this subsection we should also mention
that the electroweak corrections to the
cross sections are known to the one-loop
accuracy. For $e^+e^-$ annihilation they have been obtained
in ref.~\cite{Holl}
and for $\g\g$ collisions in ref.~\cite{Denn}.

\subsection{Forward-backward asymmetry. }
The  important parameter related to threshold production
is a space
asymmetry of the differential cross sections. This parameter gives more
detailed information on the process and allows one to obtain
independent experimental data for further testing the theory.
The forward-backward asymmetry of the top quark production
is defined as a difference of cross sections averaged over forward
and backward semispheres eith respect to the electron beam direction
devided by the total cross section.
A nonvanishing asymmetry appears in
$e^+e^-$ annihilation due to the axial coupling of the top
quark to the
$Z$-boson. The expression of this parameter for energies
near the
threshold is given by \cite{Sum}
\be
A_{FB}={E_{VA}\over
D_V}\left(1+{c^a_1-c^v_1\over 2}
{C_F\al_s\over \pi} \right)\Phi(k)
\label{afb}
\ee
where
\[
E_{VA}=q_eq_ta_ea_td+2v_ea_ev_ta_td^2
\]
is the  electroweak factor.
The expression for the asymmetry in eq.~(\ref{afb})
is given in NLO and the explicit correction of order
$\al_s$ is taken in the linear approximation that leads to the
manifest difference of axial and vector hard coefficients
in this order.

The dynamical quantity is a function
\be
\Phi(k)={1\over
m_t}{{\rm Re}\int \tilde G^*(p,k) \tilde F(p,k)
p^3dp \over \int \tilde G^*(p,k)
\tilde G(p,k)p^2dp}
\label{phidef}
\ee
that describes the overlap of the $S$ and $P$
partial waves.  Here ${\bf p}\tilde F(p,k)$ and $\tilde G(p,k)$ are the
Fourier transforms of  $i\partial_{\bf y}G({\bf x},{\bf y},k)|_{y=0}$
and $G(x,0,k)$ correspondingly.
In the Born approximation the
expression for the function $\Phi(\beta)$ can be found in the explicit
form and is rather simple $\Phi(\beta)={\rm Re}\,\beta$.
It vanishes
for the real values of energy below threshold.

\subsection{Top quark polarization.}
The longitudinal top quark polarization in  the process
$e^+e^-\rightarrow t\bar t$ averaged over the production angle
reads \cite{Har,Chi}
\[
\langle P_L\rangle =-{4\over 3}{D_{VA}\over D_V}
\left(1+{c^a_1-c^v_1\over 2}{C_F\al_s\over \pi}\right)\Phi(k)
\]
where
\[
D_{VA}=q_eq_tv_ea_td+(v_e^2+a_e^2)v_ta_td^2
\]
and $\Phi(k)$ is given by eq.~(\ref{phidef}).
This function enters also the expression for the
averaged longitudinal top quark polarization in
$\g\g\rightarrow t\bar t$ process with the same
helicity colliding photons \cite{FKK}
\[
\langle P_L\rangle =\pm 2\left(1+{c^{+-}_1-c^{++}_1\over 2}{C_F\al_s\over \pi}\right)
\Phi(k)
\]
where $+$ $(-)$ correspond to the positive (negative) helicity
photons.

The extension of the above expressions to
the general
electron/positron polarization and photon helicity
and to other component of the polarization vector can be found in
the literature (e.g. refs.~\cite{Har,FKK}).

\subsection{The effects of the finite top quark width.}
As has already been mentioned the sufficiently large
$t$-quark decay width suppresses the nonperturbative
effects of strong interactions at large $(\sim 1/\Lambda_{QCD})$
distances and makes the perturbation theory
applicable for the description of the $t$-quark threshold dynamics.
The near-threshold dynamics is nonrelativistic
and is rather insensitive to
the hard momentum details of $t$-quark
decays. Therefore as the leading order approximation 
the instability of the top quark can be
parameterized with the constant mass operator.
The finite top
quark width can then be taken into account
by the direct replacement $m_t\rightarrow m_t-i\G_t/2$
in the relevant argument $s-4m_t^2$
describing the functional dependence of physical quantities near the
threshold, or $E\rightarrow E+i\G_t$ \cite{FK}.
This approximation accounts for the leading imaginary 
electroweak contribution to the leading order NRQCD Lagrangian.  
Since the essential features of the physical situation
are caught within this approximation
we neglect the electroweak effects in  higher orders
in the strong coupling constant and heavy quark velocity. 
However, in the case of $P$ wave production and $S-P$ wave
interference the above prescription  is not sufficient
for a proper description of the entire effect of the
non-zero top quark width \cite{ttgg} and more thorough analysis
is necessary (see Sections~3.2,~3.3).

In the context of the top quark finite lifetime  
we should also  mention
the unfactorizable corrections due to the top quark
interaction with the decay products which are suppressed in the total
cross sections  \cite{FKM} but should be taken into account  
as  NLO contributions to the angular distribution
and top quark polarization \cite{PetSum}.

\section{Nonrelativistic Green function beyond the leading
order.}
The basic quantity in the analysis of the threshold effects is
the nonrelativistic Green function of the  Schr{\"o}dinger
equation~(\ref{Schr}).
The  Green function has a standard
partial wave decomposition
\be
G({\bf x},{\bf y},k)=\sum^\infty_{l=0}(2l+1)
(xy)^lP_l({\bf xy}/xy)G_l(x,y,k)
\label{lexp}
\ee
where $P_l(z)$ is a Legendre polynomial.
The partial waves of the Green function of the pure
Coulomb Schr{\"o}dinger equation
$G^C({\bf x},{\bf y},k)$ are known explicitly
\be
G^C_l(x,y,k)={m_tk\over 2\pi}(2k)^{2l}e^{-k(x+y)}
\sum_{m=0}^\infty {L_m^{2l+1}(2kx) L_m^{2l+1}(2ky)m!\over
(m+l+1-\nu)(m+2l+1)!}
\label{Gcl}
\ee
where $\nu=\lambda / k$, $\lambda =\alpha_s C_F m_t/2$
with $\al_s$ is taken at the soft scale $\mu_s$.
$L^\al_m(z)$ is a Laguerre  polynomial which is chosen
in the form
\[
L_m^\al(z)={e^zz^{-\al}\over m!}\left({d\over dz}\right)^m
(e^{-z}z^{m+\al}) .
\]
We, however, need to know the nonrelativistic Green function
for the NNLO  Hamiltonian of  the following  form
\[
{\cal H}={\cal H}_C+\Delta{\cal H}   .
\]
The second term of this expression describes the corrections 
to the  Coulomb Hamiltonian
\be
\Delta{\cal H}=-{{\bf \Delta}_{\bf x}^2\over 4m_t^3}
+\Delta_1V(x)+\Delta_2V(x)
+\Delta_{NA}V(x)+\Delta_{BF}V({\bf x},\partial_{\bf x},{\bf S}) .
\label{Ham}
\ee
The first term of the equation is the standard correction to the 
kinetic  energy operator,
$\Delta_{NA}V(x)=-C_AC_F\al_s^2/(2m_tx^2)$
is the so called non-Abelian potential
of quark-antiquark interaction \cite{Gup} and 
$\Delta_{BF}V({\bf x},
\partial_{\bf x},{\bf S})$
is a standard Breit-Fermi potential
known since long ago (only the overall
color factor $C_F$
is new).
The Breit-Fermi potential contains the quark
spin operator ${\bf S}$, {\it e.g.}~\cite{Landau}.
In NNLO the cross section $R^{v}$ is saturated by
the final state configuration of $t\bar t$ pair
with $l=0$, $S=1$
while $R^{++}$ cross section
is saturated by $l=0$, $S=0$ configuration.
The Breit-Fermi potential takes the following form
when considered on the $l=0$ states
\[
\Delta_{BF}V(x)={C_F\al_s\over x}{{\bf \Delta_x}\over m_t^2}+
A^i{C_F\al_s\pi\over m_t^2}\delta({\bf x})
\]
where $A^v=11/3$ corresponds to the spin one final state of
$e^+e^-\rightarrow t\bar t$ production and $A^{++}=1$
corresponds to the spin zero final state of $\g\g\rightarrow t\bar t$
production.

The terms $\Delta_iV$ ($i=1,2$) represent
the first and second order perturbative QCD corrections to the
Coulomb potential \cite{Fish,Peter}
\[
\Delta_1 V(x)={\al_s\over 4\pi}V_C(x)(C_0^1+C_1^1\ln(x\mu_s)),
\]
\[
\Delta_2 V(x)
=
\left({\al_s\over 4\pi}\right)^2V_C(x)(C_0^2+C_1^2\ln(x\mu_s)
+C_2^2 \ln^2(x\mu_s))
\]
where
\[
C_0^1=a_1+2\beta_0\gamma_E,\qquad C_1^1=2\beta_0,
\]
\[
C_0^2=\left({\pi^2\over 3}+4\gamma_E^2\right)\beta_0^2
+2(\beta_1+2\beta_0a_1)\gamma_E+a_2,
\]
\[
C_1^2=2(\beta_1+2\beta_0a_1)+8\beta_0^2\gamma_E,
\qquad
C_2^2=4\beta_0^2,
\]
\[
a_1={31\over 9}C_A-{20\over 9}T_Fn_f,
\]
\[
a_2= \left({4343\over 162}+4\pi^2-{\pi^4\over 4}
+{22\over3}\zeta(3)\right)C_A^2-
\left({1798\over 81} + {56\over 3}\zeta(3)\right)C_AT_Fn_f
\]
\[
-\left({55\over 3} - 16\zeta(3)\right)C_FT_Fn_f
+\left({20\over 9}T_Fn_f\right)^2,
\]
\[
\beta_1={34\over 3}C_A^2-{20\over 3}C_AT_Fn_f-4C_FT_Fn_f.
\]
Here $\al_s$ is defined in $\overline{\rm MS}$
renormalization scheme. The invariants
of the color symmetry
$SU(3)$ group have the following numerical values
for QCD: $C_A=3$,
$C_F=4/3$, $T_F=1/2$, $n_f=5$ is the number of light quark
flavors, $\beta_0=11C_A/3-4T_Fn_f/3$ is the first $\beta$-function
coefficient, $\gamma_E=0.577216\ldots$ is the Euler constant
and $\zeta(z)$ is the Riemann $\zeta$-function.
The solution to eq.~(\ref{Schr}) with the Hamiltonian~(\ref{Ham}) can
be found within the standard nonrelativistic perturbation theory around
the Coulomb Green function as a leading order approximation
\[
G({\bf
x},{\bf y},k)
=G_C({\bf x},{\bf y},k)+\Delta G({\bf x},{\bf y},k),
\]
\be
\Delta G({\bf x},{\bf y},k)=-\int G_C({\bf x},{\bf z},k)
\,\Delta{\cal H }\,G_C({\bf z},{\bf y},k)d{\bf z}+\ldots
\label{totcorr}
\ee
In the previous section the threshold effects
in the basic observables were reduced to
three universal functions: the Green function at the origin
which is saturated by the $S$ wave contribution, the derivative
of the Green function at the origin
which is saturated by the $P$ wave contribution and the function
$\Phi(k)$ which describes the $S-P$ wave interference.
These functions are analyzed in detail in Sections 3.1-3.3

\subsection{$S$ wave production.}
Only the $l=0$ component of the Green function (\ref{lexp})
contributes
to its value at the origin
\[
G(0,0,k)=G_0(0,0,k).
\]
The explicit
expression for the Coulomb part of the Green function has the form
\[
G^C_0(x,0,k)|_{x\rightarrow 0}=
{m_t\over 4\pi}\left({1\over x}-
2\lambda\ln\left({2x\mu_f}\right)
-2\lambda\left({k\over 2\lambda}+
\ln\left({k\over \mu_f}\right)
\right.\right.
\]
\be
+2\gamma_E-1+
\Psi_1\left(1- \nu\right)\bigg)\bigg)
\label{G0}
\ee
where $\Psi_n(z)=d^n\ln{\Gamma(z)}/dz^n$ and
$\Gamma(z)$ is the Euler $\Gamma$-function.
The energy independent finite part of this expression is chosen for
later convenience.
Eq.~(\ref{G0}) can be most easily obtained
from the general expression for the Coulomb partial waves
\be
G^C_l(x,0,k)=
{m_tk\over 2\pi}(2k)^{2l}e^{-kx}\Gamma(l+1-\nu)U(l+1-\nu,2l+2,2kx)
\label{hyp}
\ee
where $U(a,b,z)$ is the confluent hypergeometric function.
In the short distance limit $x\rightarrow 0$
the Coulomb Green function
$G^C({\bf x},0,k)=G^C_0({x},0,k)$ has $1/x$ and $\ln(x)$
divergent terms. These terms, however, are energy independent
and do not contribute
to the cross section.
Hence these terms can be
subtracted without affecting any physical results.
The quantity $\mu_f$ in eq.~(\ref{G0}) is an
auxiliary parameter, the factorization scale,
which drops out from the physical observables.

The NLO correction $\Delta_1G$ to eq.~(\ref{G0})
due to the first iteration
of $\Delta_1V$ term of the QCD potential
has been found in ref.~\cite{KPP}
where the simple and efficient framework for computation
of higher orders was formulated.
The result of the evaluation of the NLO correction is
\[
\Delta_1G_0(0,0,k)
={\al_s\beta_0\over 2\pi}{\lambda m_t\over 2\pi}\left(
\sum_{m=0}^\infty F(m)^2(m+1)
\left(L_1(k)+\Psi_1(m+2)\right)-2\sum_{m=1}^\infty\sum_{n=0}^{m-1}
F(m)
\right.
\]
\[
\left. \times F(n){n+1\over m-n} +2\sum_{m=0}^\infty F(m)
\left(L_1(k) - 2\gamma_E-\Psi_1(m+1)\right)
-\gamma_E L_1(k)+{1\over 2}L_1(k)^2
\right)
\]
where
\[
L_1(k)=\ln\left({\mu_se^{C_0^1/C_1^1}\over 2k }\right)
\]
and
\[
F(m)={\nu\over (m+1)\left(m+1-\nu\right)}.
\]
The NNLO correction $\Delta_2^{(2)}G$
due to  $\Delta_2V$ part of the potential
and the correction $\Delta_2^{(1)}G$ due to the
second iteration of $\Delta_1V$ part of the correction to the Coulomb
static potential  have been obtained in refs.~\cite{KPP,PP}.
While the technique is rather straightforward the results
of the calculations are cumbersome and explicit formulae are
relegated to Appendix~A.

The method of calculation of the correction to the Green
function at the origin due to logarithmic terms in the potential
is described in details in ref.~\cite{PP1}.
It is based on the representation of the Coulomb Green function as an
expansion over the Laguerre polynomials~(\ref{Gcl}).
This representation is very close to the standard physical
expansion over the eigenfunctions that makes the technique transparent
and easily interpretable in physical terms.
It is equally suitable for any partial wave contribution
as has been shown in
ref.~\cite{PP2}
where results for the $P$ wave production were found.
The results for the $S$ wave part of the corrections
were reproduced
within a different technical framework based on an integral
representation of the Coulomb Green function in ref.~\cite{MY}.

The corrections to the Coulomb Green function at the origin
due to ${\bf\Delta}^2$, $V_{NA}$ and  $V_{BF}$ terms have been
presented in \cite{Hoang,Mel}.
They are of the following explicit form
\[
\Delta_{{\bf \Delta}^2,NA,BF}G=
{m_t\over 4\pi}{k^2\over m_t^2}\left(
{5\over 8}k+
4\lambda\left(
\ln\left({k\over \mu_f}\right)+\gamma_E
+\Psi_1\left(1- \nu\right)\right)
-{11\over 8}C_F\al_s\nu
\Psi_2\left(1-\nu\right)\right)
\]
\be
+\pi{C_F\al_s\over m_t^2}
\left(5-A^i+2{C_A\over C_F}\right)G_C(0,0,k)^2 \, .
\label{BFcorr}
\ee
In the course of evaluation of this correction
to the nonrelativistic Green function one encounters
the ultraviolet divergence in the imaginary part of the
Green function contained
in the last term of eq.~(\ref{BFcorr}). This divergence
is related to the singular behavior of the
Coulomb Green function at the origin.
The particular form of this divergence
depends on the regularization
procedure. The divergence appears in the process of scale separation
and is a consequence of the fact that the nonrelativistic
approximation is not adequate for the description of the
short distance effects. 
The  hard coefficient $C^{v,++}$ computed within the same regularization
procedure as the Green function itself
must have the infrared singular
term which exactly cancels the one appearing in the Green function.
The hard coefficient can be evaluated  by matching the effective  
and full theory cross sections in the weak coupling limit \cite{H1,Hoang}
or by an explicit splitting of the hard and soft contributions
using, for example, the scale factorization  in  
dimensional regularization \cite{BSS,PinSot1,CMY}.  
Let us consider the cancelation of the divergences and 
determination of the hard coefficient in the matching scheme. 
The natural regularization  for the analysis of  
the  hard part of the corrections is the  dimensional one \cite{c1,c2}. 
In $4-2\ve$ dimensions the infrared divergence of the hard contribution 
in NNLO has the form of the first order pole in $\ve$.
The Coulomb Green function at the origin in
eq.~(\ref{BFcorr}) can be regularized in the same way. 
The dimensionally regularized Coulomb Green function at the origin
can be defined as follows (see Appendix~B)
\be G^{d.r.}_C(0,0,k)=-{m_t\over 4\pi}\left(k +
2\lambda \left(-{1\over 2\ve}+\ln\left({k\over\mu_f}\right)
+\gamma_E+\Psi_1(1-\nu)\right)\right) +O(\ve).
\label{dimreg}
\ee
Note that the Green function regularized in this 
way differs from that which has been obtained in 
ref.~\cite{CMY} within another scheme
of dimensional regularization. 
In contrast to eq.~(\ref{G0})
in~eq.~(\ref{dimreg}) there is no divergence in the Born
approximation. The Green function in this approximation is a
nonrelativistic free propagator and is proportional to
$k$. The first order pole in $\ve$  appears
only in the first order in $\al_s$. 
The  $O(\al^2)$ singular $1/\ve$ term
in the imaginary part of eq.~(\ref{BFcorr}) is proportional 
to ${\rm Im}(G_C(0,0,k))$ and, therefore,  can be absorbed  by
the redefinition of the hard  coefficient $C^{v,++}$. 
For the  Green function this redefinition
results in the   substitution
$G^{d.r.}_C(0,0,k)\rightarrow G^s_C(0,0,k)$
in eq.~(\ref{BFcorr}) where
the ``subtracted'' Green function reads
\be
G^s_C(0,0,k)=G^{d.r.}_C(0,0,k)-{m_t\lambda\over 4\pi} {1\over
\ve}. 
\label{G0r}
\ee
Within the redefined hard coefficient    
the  $O(\al^2)$ 
``ultraviolet'' $1/\ve$ term stemming from the corrections to the 
Green function eq.~(\ref{BFcorr}) exactly cancels the $O(\al^2)$ 
``infrared'' $1/\ve$ term of the hard  part of the corrections
regularized in the same way. 
Then  the (finite) coefficient 
$C^{v,++}$ can be found directly by matching the 
effective theory expression
for the cross sections and the result of perturbative QCD calculation
of the spectral density in the formal limit $\al_s\ll \bt\ll 1$ up to
the order $\al_s^2$ for $\mu_h = \mu_s$.
Eqs.~(\ref{Ree},~\ref{Rgg}) in the matching limit take the form
\[
R^v={3\over
2}N_cq_t^2\beta\left(\left(1+\left(1-B^v\right) \bt^2\right)+
C_F{\al_s\over \pi}\left({\pi^2\over 2}{1\over\bt}+ c^v_1+ {\pi^2\over
3}\bt\right)\right.
\]
\[
+C_F\left({\al_s\over \pi}\right)^2
\left(C_F{\pi^4\over 12}{1\over\bt^2}+
\pi^2\left(C_F{c^v_1\over 2}+{1\over 8}
\left(-C_1^1\ln{\left({2\bt m_t\over\mu_s}\right)}+C_0^1-2\bt_0\g_E\right)
\right){1\over\bt}\right.
\]
\be
\left.
+C_F{5\pi^4\over 36}+c^v_2-C_F\pi^2\left({5-A^v\over 2}+{C_A\over C_F}
\right)\ln{\left({\bt m_t\over\mu_f}\right)}\right),
\label{re}
\ee
\[
R^{++}=6N_cq_t^4\beta
\left(\left(1+\left(1-B^{++}\right)\bt^2\right)+
C_F{\al_s\over \pi}\left({\pi^2\over 2}{1\over\bt}+ c^{++}_1+
{\pi^2\over 3}\bt\right)\right.
\]
\[
+C_F\left({\al_s\over \pi}\right)^2
\left(C_F{\pi^4\over 12}{1\over\bt^2}+\pi^2
\left(C_F{c^{++}_1\over 2}+{1\over 8}
\left(-C_1^1\ln{\left({2\bt m_t\over\mu_s}\right)}+C_0^1-2\bt_0\g_E\right)
\right){1\over\bt}\right.
\]
\be
\left.
+C_F{5\pi^4\over 36}+c^{++}_2-C_F\pi^2\left({5-A^{++}\over 2}+
{C_A\over C_F}
\right)\ln{\left({\bt m_t\over\mu_f}\right)}\right)
+R^{++}_P 
\label{rg}
\ee
where the terms of the relative order $O(\bt^2)$ are kept.
In eq.~(\ref{rg}) the $P$ wave contribution
\[
R^{++}_P=6N_cq_t^4\beta^3+\ldots 
\]
is due to the derivative term in  eq.~(\ref{Rpp})
to be discussed.
By comparing eq.~(\ref{re}) with the NNLO QCD
result for the cross section $R^v$ expanded in the
velocity of the heavy quark near the threshold \cite{c1,sd}
one finds
\be
c^v_2=\tilde c^v_2-c^v_1{\beta_0\over 2}
\ln{\left(m_t\over\mu_h\right)}+\pi^2\left({2\over 3}C_F+C_A
\right)\ln{\left(m_t\over\mu_f\right)},
\label{ce0}
\ee
where the coefficient $\tilde c^v_2$  has been obtained in 
refs.~\cite{Hoang,Mel}
\[
\tilde c^v_2=\left({39\over 4}-\zeta(3)+{4\pi^2\over 3}
\ln{2}-{35\pi^2\over 18}\right)C_F-\left({151\over 36}+{13\over 2}
\zeta(3)+{8\pi^2\over 3} \ln{2}-{179\pi^2\over 72}\right)C_A
\]
\[
+\left({44\over 9}-{4\pi^2\over 9}+{11\over 9}n_f\right)T_F .
\]
The first logarithm in eq.~(\ref{ce0}) is determined
by the renormalization group running of the strong coupling constant
in the hard momentum regime
and is proportional to the first coefficient
of the $\beta$ function.
Thus, both the hard  coefficient and 
the imaginary part of the Green function do not depend
on the  normalization point of $\al_s$ in the fixed 
order of perturbation theory so one can use   
different scales for $\al_s$ in these quantities. 
The second logarithm corresponds to the anomalous dimension of the hard  
coefficient                  
and precisely cancels the factorization scale dependence of the Green
function due to eq.~(\ref{BFcorr}) making the total
result  independent of the factorization scale.
Note that the use of different hard and soft 
normalization scales leads to the incomplete cancelation
of the factorization scale dependence which, however,
is 
the higher order  
($O(\al_s^3)$) effect.  

As it has been mentioned above  one can bypass  the direct matching 
by the consistent use of the same subtraction scheme within
dimensional regularization for both hard coefficient and Green function.
Then matching is 
automatic \cite{BenSmi,PinSot1,CMY}.   
In this approach the hard coefficient
is completely determined by the hard renormalization
coefficient of the nonrelativistic vector current \cite{BSS}. 
However, in order to compute the corrections to the Green function 
in this case one has to define accurately the 
Breit-Fermi Hamiltonian and the Green function 
in $3-2\ve$ dimensions \cite{BSS,CMY}
(in our analysis we use  three dimensional  Breit-Fermi 
Hamiltonian and direct regularization of the Green function  
therefore the matching is necessary in order to fix the constant 
relating the two regularization schemes).

The NNLO  analysis of the $R^{++}$ cross section is still absent
and the constant in the hard coefficient is unknown.
The logarithmic part of the NNLO
contribution to $C^{++}(\al_s)$ reads
\be
c^{++}_2=\tilde c^{++}_2- c^{++}_1{\beta_0\over 2}
\ln{\left(m_t\over\mu_h\right)}+\pi^2\left(2C_F+C_A
\right)\ln{\left(m_t\over\mu_f\right)}
\label{cg}
\ee
where $\tilde c^{++}_2$ is a constant to be determined.
The relativistic correction to this cross section,
however, can be extracted from the calculations presented earlier in
the literature. Comparing the known result \cite{INOK}
\be
R^{++}(\beta)=6q_t^4N_c\beta\left(1+{2\over
3}\beta^2+O(\beta^4)\right)
\label{rel}
\ee
with our expressions from
eqs.~(\ref{rg},~\ref{rel}) we find
\[
B^{++}={4\over3}.
\]
The
Green function at the origin can be written in the form which includes
only single poles in the energy variable. This form seems to be natural
for a Green function of the  nonrelativistic Schr{\"o}dinger equation
\be
G(0,0,E)=\sum_{m=0}^\infty{|\psi_{0m}(0)|^2\over E_{0m}-E}+
{1\over \pi}\int_0^\infty{|\psi_{0E'}(0)|^2\over E'-E}
dE'
\label{endenom}
\ee
where $\psi_{0m,E'}(0)$ is the wave function at the origin,
the sum goes over the bound states
and the integral goes over the states of a continuous part of the
spectrum.
In this way the corrections to the Green function stemming from
the discrete part of the spectrum
reduce to
corrections to Coulomb bound state energy levels
\[
E_{0m}=E^C_{0m}\left(1+\Delta_1E_{0m}+\Delta_2E_{0m}\right)
\]
and to the values of Coulomb bound state wave functions at the origin
\[
|\psi_{0m}(0)|^2=|\psi^C_{0m}(0)|^2\left(1+\Delta_1\psi^2_{0m}
+\Delta_2\psi^2_{0m}\right)
\]
where
\[
E^C_{0m}=-{\lambda^2\over m_t(m+1)^2}, \qquad
|\psi^C_{0m}(0)|^2={\lambda^3\over \pi(m+1)^3},
\]
\[
\Delta_2E_{0m}=\Delta_{{\bf \Delta}^2,NA,BF}E_{0m}
+\Delta^{(2)}_2E_{0m} +\Delta^{(1)}_2E_{0m},
\]
\[
\Delta_2\psi^2_{0m}=\Delta_{k^2}\psi^2_{0m}+
\Delta_{{\bf\Delta}^2,NA,BF}\psi^2_{0m}+\Delta^{(2)}_2\psi^2_{0m}
+\Delta^{(1)}_2\psi^2_{0m}
\]
and $\Delta_{k^2}\psi^2_{0m}$ is the correction due to
relativistic corrections parameterized by the coefficients $B^i$
which we include into the definition of the wave function.

In NLO an explicit analytical expression for the corrections
to the bound state parameters has the form \cite{MY,PP1,PY}
\[
\Delta_1E_{0m}={\alpha_s\beta_0\over \pi}\left(\bar L_1(m)+\Psi_1(m+2)\right),
\]
\[
\Delta_1\psi^2_{0m}={\alpha_s\beta_0\over 2\pi}
\left(3\bar L_1(m)+\Psi_1(m+2)-2(m+1)\Psi_2(m+1)
-1-2\gamma_E+{2\over m+1}
\right)
\]
where $\bar L_1(m)=L_1(\lambda /(m+1))$.
The expressions of the NNLO
corrections  to the energy levels \cite{MY,PP1,PY} and
wave functions at the origin \cite{MY,PP1}
are rather cumbersome and given in Appendix~C, D.

The continuum contributions in eq.~(\ref{endenom})
can be directly found by subtracting the
discrete part of these equations expanded around  the Coulomb
approximation up to NNLO
\[
\sum_{m=0}^\infty{|\psi^C_{0m}(0)|^2\over E^C_{0m}-E}\left(
1+\Delta_1\psi^2_{0m}+\Delta_2\psi^2_{0m}
+{\left(1+\Delta_1\psi^2_{0m}\right)\Delta_1E_{0m}+\Delta_2E_{0m}\over
1-E/E^C_{0m}}
\right.
\]
\[
\left.
+{\Delta_1E_{0m}^2\over (1-E/E^C_{0m})^2}\right)
\]
from the result obtained within the
nonrelativistic perturbation theory for the Green function
at the origin~(\ref{totcorr}) multiplied by $(1-B^iE/m_t)$.
This procedure removes the double and triple poles from
eq.~(\ref{totcorr}) and leaves only
the single poles in the expression for
the Green function~(\ref{endenom}).

An important consequence of the relatively large
top quark width is that the most of Coulomb resonances are smoothed
out. The numerical analysis shows that only the ground state
resonance in the cross sections
is distinguishable.
Its separation from others is not completely covered
by the infrared cutoff provided by the top quark width.
Indeed, using the pure Coulomb formulas for estimates within the order of
magnitude we find
\[
|E^C_{00}-E^C_{01}|={3\lambda^2\over 4m_t}\approx 0.6~{\rm GeV}
\]
to be compared with the top quark width $\G_t=1.43~{\rm GeV}$.
The second spacing between radial excitations for the $l=0$ partial wave
and the first spacing for the $l=1$ partial wave  are, however, much smaller
\[
|E^C_{01}-E^C_{02}|=|E^C_{10}-E^C_{11}|=
\frac{5}{36}{\lambda^2\over m_t}\approx
0.11~{\rm GeV} \, .
\]
Therefore the contributions of higher resonances are
completely smeared out with the top quark width.

In  the  limit of vanishing top quark width
the NNLO approximation for the energy
of the first resonance in $e^+e^-$ annihilation reads
\[
E_{00}^v=-{\lambda^2\over m_t}\left(1+{\alpha_s\over 4\pi}
2C_1^1\left(L_1(\lambda)+1-\gamma_E\right)
+\left({\alpha_s\over 4\pi}\right)^2\bigg(2C_1^2(L_2(\lambda)+1-\gamma_E)
\right.
\]
\[
\left.
+(C_1^1)^2\left((L_1(\lambda)-\gamma_E)^2+1-{\pi^2\over 3}-\Psi_3(1)\right)
+2C_2^2\left((L(\lambda)+1-\gamma_E)^2-1+{\pi^2\over 6}\right)
\right)
\]
\be
\left.
+C_F^2\al_s^2\left({C_A\over C_F}+{1\over 48}\right)
\right)
\label{e2}
\ee
where
\[
L(\lambda)=\ln\left({\mu_s\over 2\lambda }\right),
\qquad L_2(\lambda)=\ln\left({\mu_se^{C_0^2/C_1^2}\over 2\lambda }\right).
\]
This value is related to the energy of the
resonance of the top quark production in  $\g\g$ collision
by the  hyperfine splitting
\[
E_{00}^v-E_{00}^{++}=
{4\over 3}{\lambda^2\over m_t}C_F^2\al_s^2\ .
\]
The convergence of the perturbation theory 
series~(\ref{e2}) is not fast.
For some typical values of the soft normalization scale  
the series for the resonance energy reads  
\[
E^v_{00}= E^{LO}_{00}(1+0.36+0.30),\qquad  \mu_s=25~{\rm GeV},
\]
\[
E^v_{00}= E^{LO}_{00}(1+0.58+0.38), \qquad \mu_s=50~{\rm GeV},
\]
\be
E^v_{00}= E^{LO}_{00}(1+0.68+0.45),\qquad \mu_s=75~{\rm GeV},
\label{div}
\ee 
\[
E^v_{00}= E^{LO}_{00}(1+0.74+0.50),\qquad \mu_s=100~{\rm GeV}.
\]
The poor convergence of the series  for the resonance
energy  can be  assigned 
to high infrared sensitivity of the pole mass (see, for example,
\cite{BB}).
The convergence can be  manifestly improved by removing the pole mass
from the theoretical expressions in favor of some less 
infrared sensitive mass parameter, for example, the short-distance \cite{MY}, 
potential-subtracted \cite{BSS} or $1S$ mass \cite{H2}.  
Note that in a finite order of the expansion all the 
mass definitions are perturbatively equivalent.  The  
infrared safe mass parameters, however, 
are ``closer'' to the physical observables since 
in contrast to the pole mass the corresponding 
perturbative  series are supposed to be 
better convergent (less divergent). 
 
Due to the finite top quark width the 
location of the  peak  (maximum) of the cross section  is not given only 
by  the position
of the ground state resonance but is also 
affected by the contribution
of the higher (smeared out) resonances and the continuum
contribution.
Due to this effect  the absolute value
of the NNLO peak energy~(\ref{e2}) counted from the threshold
is less than the absolute value of the energy of the ground state 
resonance $E^{v,++}_{00}$
by about $200~{MeV}$ {\it i.e.} $\sim 7\%$. This shift is essentially smaller
than the one related to the perturbative QCD 
corrections to Coulomb values but considerably larger than the 
leading nonperturbative contribution
due to the gluon condensate \cite{VL} which is suppressed parametrically 
as $(\Lambda_{QCD}/\lambda)^4<1\%$. 

\subsection{$P$ wave production.}
The derivative of the Green function at the origin
is saturated with its $l=1$ component
and explicitly given by the relation
\[
\partial^2_{\bf x\bf y}
G({\bf x},{\bf y},k)=
9G_1(0,0,k) .
\]
For the Coulomb Green function from eq.~(\ref{hyp}) we obtain
the closed formula for the partial wave $l=1$ Green function 
at the small space separation of particle
\[
G^C_1(x,0,k)|_{x\rightarrow 0}=
{m_t\over 36\pi}
\left({3\over x^3}+{3\lambda\over x^2}+
{6\lambda^2-3k^2\over 2x}
+2\lambda(k^2-\lambda^2)\ln(2x\tilde \mu_f)
\right.
\]
\be
\left.
+\lambda\left(2(k^2-\lambda^2)\left({k\over 2\lambda}+
\ln\left({k\over \tilde\mu_f}\right)
+2\gamma_E-{11\over 6}+
\Psi_1\left(1- \nu\right)\right)+{k^2\over 2}\right)\right)
\label{dG0}
\ee
where $\tilde \mu_f$ is the analog of the parameter $\mu_f$ for
the $l=0$ partial wave.
In the short distance limit $x\rightarrow 0$
the derivative of the Coulomb Green function
(or the partial wave with $l=1$) has $1/x^n$ $(n=1,2,3)$
and $\ln(x)$ singularities.
In contrast to the case of the
$S$ wave production, the value
at the origin for the $P$ wave partial Green function
contains divergent terms that explicitly 
depends on the energy (or wave vector)
$k$. 
However, these terms do not contribute to the cross section for 
the vanishing top quark width 
$\G_t=0$ because they have no 
discontinuity across the physical cut in the complex plane of the 
energy variable in the approximation of top quark zero width. The case 
of the non-zero top quark width requires 
to perform a more detailed analysis given 
below.

The correction to the $l=1$ partial wave at the origin
due to the first iteration of $\Delta_1V$ term of the QCD potential
has been found in ref.~\cite{PP2}
\[
\Delta_1G_1(0,0,k)
={\al_s\beta_0\over 2\pi}{\lambda m_tk^2\over 18\pi}\left(
\sum_{m=0}^\infty \tilde F(m)^2(m+1)(m+2)(m+3)
\left(L_1(k)+\Psi_1(m+4)\right)
\right.
\]
\[
-2\sum_{m=1}^\infty\sum_{n=0}^{m-1}
\tilde F(m)\tilde F(n){(n+1)(n+2)(n+3)\over m-n} +2\sum_{m=0}^\infty
\tilde F(m)\bigg(2\tilde J_0(m)+(m+1)(m+2)L_1(k)
\]
\[
\left.
+(1+\nu)(\tilde J_1(m)+(m+1)L_1(k))
+{\nu(\nu+1)\over 2}(\tilde J_2(m)+
2L_1(k))\bigg)
+\tilde I(k) \right)
\]
where
\[
\tilde F(m)={\nu(\nu^2-1)\over (m+2-\nu)(m+1)(m+2)(m+3)}
\]
\[
\tilde J_0(m)=-2\Psi_1(m+1)-4\gamma_E +3,
\]
\[
\tilde J_1(m)=(m+1)(-\Psi_1(m+2)-2\gamma_E +2),
\]
\[
\tilde J_2(m)={(m+1)(m+2)\over 2}\left(-\Psi_1(m+3)-
2\gamma_E +{3\over2}\right) ,
\]
\[
\tilde I(k)=-{(\gamma_E-1)^2\over 2}-{\pi^2\over 12}
-(4-3\gamma_E)\nu +{1-9\gamma_E+6\gamma_E^2+\pi^2\over 4}\nu^2+
{1-3\gamma_E\over 2}\nu^3+{1-\gamma_E\over 4}\nu^4
\]
\[
+\left(\gamma_E-1-3\nu+{9-12\gamma_E\over 4}\nu^2
+{3\over 2}\nu^3+{1\over 4}\nu^4\right)L_1(k)
+\left(-{1\over 2}+{3\over 2}\nu^2\right)L_1(k)^2
\]
For the derivative of the Green function at the origin
(or for the $l=1$ partial wave)
the analog of eq.~(\ref{endenom}) reads
\[
\partial^2_{\bf xy}G({\bf x},{\bf y},E)=
\sum_{m=0}^\infty{|\psi'_{1m}(0)|^2\over E_{1m}-E-i0}+
{1\over \pi}\int_0^\infty{|\psi'_{1E'}(0)|^2\over E'-E-i0}
dE'
\]
where, symbolically,
\[
|\psi'_{1m,E'}(0)|^2=
\partial_{\bf x}\psi^*_{m,E'}({\bf x})
\partial_{\bf y}\psi_{m,E'}({\bf y})|_{x,y=0}.
\]
Here $E_{1m}$ is the $l=1$ bound state energy.
In NLO approximation these quantities read \cite{PP2}
\[
E_{1m}=-{\lambda^2\over m_t(m+2)^2}\left(1+
{\alpha_s\over 4\pi}
2C_1^1\left(\bar L_1(m+1)+\Psi_1(m+4)\right)\right)\ ,
\]
and
\[
|\psi'_{1m}(0)|^2={\lambda^5\over \pi}{(m+1)(m+3)\over (m+2)^5}
\left(1+
{\alpha_s\over 4\pi}C_1^1\left(5\bar L_1(m+1)+5\Psi_1(m+4)-
{\pi^2\over 3}(m+2)-1
\right.\right.
\]
\[
\left.\left.
+2\sum^{m-1}_{n=0}{(n+1)(n+2)(n+3)\over
(m+1)(m+3)(m-n)^2}\right)\right)\ .
\]
The continuum contribution is obtained in the same way
as it was done in the previous section for the $S$
wave production.

In the case of $P$ wave production the simple 
shift $E\rightarrow E+i\G$
in the  nonrelativistic approximation is not sufficient
to describe properly the entire effect of the non-zero top quark
width \cite{ttgg}.
Indeed, eq.~(\ref{dG0}) in the limit $x\rightarrow 0$
with
the nonvanishing width
has the divergent imaginary part
with the leading power singularity $\sim\G_t/x$ related to the
free Green function singularity and
the logarithmic singularity $\sim\G_t\al_s\ln(x)$
produced by the one Coulomb gluon exchange.
The presence of these singularities
clearly indicates that the coefficient
of the constant term linear in $\G_t$
gets a contribution from the large momentum region and cannot
be obtained within the pure nonrelativistic  approximation.
Like the hard coefficients it should be computed
in  relativistic theory.
Parametrically this contribution is not suppressed
in comparison to the pure nonrelativistic
contribution in the threshold region. At $E=0$, for example,
the ratio between the relativistic (proportional to $\G_t$)
and nonrelativistic (Coulomb) contributions is of
order ${\G_t\over\al_s^2m_t}\sim 1$. Since we are
interested in the NLO corrections the  $O(\G_t\al_s)$
term also has to be taken into account.
By construction, the
nonrelativistic effective theory has to reproduce the perturbation
theory predictions in the formal matching 
limit 
$\alpha_s,~ \bt \ll \Gamma_t/m_t\ll 1$
where both effective theory and perturbation theory
descriptions are valid.
Thus one has to compute $O(\G_t)$ and $O(\G_t\al_s)$
terms within the relativistic perturbation theory and then to fix the
parameters of the effective nonrelativistic theory so that it
reproduces the perturbative results in the matching
limit. Within the relativistic perturbation theory  
the relevant contributions can be obtained   by inserting 
the complex momentum-dependent  mass
operator into the top quark propagator at $\bt =0$ (only
the leading terms in $\Gamma_t/m_t$ should be kept). 
In the leading order in $\al_s$
this procedure has been done in refs.~\cite{ttgg}.
The result reads
\[
G^C_1(0,0,k)|_{\G_t}= {m_t^3\over 36\pi}g_1\G_t
\]
where
$g_1$ is a coefficient coming from the
relativistic treatment with numerical value
$g_1=0.185\ldots$
For the term of the order $O(\G_t\al_s)$
the necessary
calculation has been performed in
ref.~\cite{PP2}. It has been
shown that the proper relativistic
analysis leads to fixing the auxiliary parameter of
eq.~(\ref{dG0}) $\tilde \mu_f=g_2 m_t$
where $g_2$ is the coefficient coming from the
relativistic treatment. Its numerical value
is $g_2=0.13\ldots$. 

Here we should note also the problem of the previous 
numerical analysis of the $P$ wave contribution \cite{KuhTeu}.
While solving the Schr{\"o}dinger equation~(\ref{Schr}) 
numerically for the finite top quark width one has to introduce an explicit
ultraviolet cutoff for the nonrelativistic expressions 
divergent in the large momentum region.
To get rid of the cutoff  dependence one has to compute
the hard contribution within the relativistic approximation
using the similar prescription for the infrared cutoff.
This, however, has not been done and 
as a consequence the $O(\G_t)$ and  $O(\G_t\al_s)$
contributions to the cross section
were not  determined within the numerical framework of 
ref.~\cite{KuhTeu}\footnote{Recently  the $O(\G_t)$ contribution has been 
estimated within the numerical approach \cite{HoaTeu} by using
the physical (relativistic) phase space for the unstable top quark
to regularize the divergence of the nonrelativistic approximation.}.
On the other hand the total $O(\G_t)$ contribution to the 
cross section is numerically small in comparison with that of the 
regular completely nonrelativistic terms of eq.~(\ref{dG0}) which 
saturate the total result for 
the energies not far below the threshold.

\subsection{$S-P$ interference.}
In the zero width approximation the function~(\ref{phidef})
allows for the following decomposition
\[
\Phi(E) =\Phi_{pol}(E)+\Phi_{con}(E)
\]
where $\Phi_{con}$ and $\Phi_{pol}$ are the continuum and
bound state poles contributions correspondingly.
This is known \cite{FKK,Chi} that the continuum contribution
is not affected by the Coulomb effects and above the threshold
one has the Born approximation result
\[
\Phi^C_{con}(E)={\rm Re}\sqrt{E\over m_t}
\]
even for the Coulomb Green function in eq.~(\ref{phidef}).
Below the threshold in the Coulomb approximation one  gets
\be
\Phi^C_{pol}(E)=\left(\sum^\infty_{m=0}{\phi_m^C\over (E^C_{1m}-E)^2}\right)
\left(\sum^\infty_{m=0}{|\psi_m(0)|^2\over (E^C_{0m}-E)^2}\right)^{-1}
\label{ph0}
\ee
where the quantities 
\[
\phi_m^C={\lambda^4\over m_t\pi}{(m+1)(m+3)\over (m+2)^5}
\]
measure  the overlap of the $S$ and $P$ wave
functions. Note that in the zero width  limit
the function $\Phi^C_{pol}$ does not vanish due to
the Coulomb degeneration of the energy levels with different $l$:
$E^C_{0m+1}=E^C_{1m}$.
It was pointed in ref.~\cite{Chi} that the continuum contribution
gets no soft corrections. Thus in NLO for a finite  top quark width
we have the simple result
\[
\Phi_{con}(E)={\rm Re}\sqrt{E+i\G_t\over m_t} .
\]
The corrections  to the pole contribution are less trivial.
They can be computed using the powerful technique developed in
refs.~\cite{KPP,PP1,PP2}. The result reads
\be
\Phi_{pol}(E)={\rm Re}\left(\sum^\infty_{m=0}{\phi_m\over
(E_{0(m+1)}-E+i\G_t)(E_{1m}-E-i\G_t)}\right)
\left(\sum^\infty_{m=0}{|\psi_m(0)|^2\over (E_{0m}-E)^2+\G_t^2}\right)^{-1}
\label{ph1}
\ee
where
\[
\phi_m=\phi_m^C\left(1+
{\alpha_s\beta_0\over \pi}\left( 4L(m+1)+4\Psi(m+4)+
{m\over m+3}-2-{\pi\over 6}(m+2)
\right.\right.
\]
\[
\left.\left.
+\sum_{n=0}^{m-1}{(n+1)(n+2)(n+3)\over (m+1)(m+3)(m-n)^2}\right)\right)
\]
In eq.~(\ref{ph1}) we keep the finite top quark width
to get a nonvanishing result since the Coulomb
degeneration is lifted by the
logarithmic corrections to the potential. Strictly speaking,
our approach is valid only if the level splitting $E_{0m+1}-E_{1m}$
is much
smaller
than the top quark width (which is
realized for the actual
values of these quantities).
For $E_{0m+1}-E_{1m}>\G_t$
the nonrelativistic analysis is not applicable
for the $S-P$ interference below the threshold
because the double poles of eq.~(\ref{ph0}) disappear and  the
nonrelativistic contribution is not enhanced in comparison
to the relativistic one in this case.

Note that for  the finite top quark width  the interference
of the free $l=0$ and $l=1$ partial waves results in the logarithmically
divergent $O(\G_t)$  term  in the numerator of  eq.~(\ref{phidef})
(this term does not include the factor $\G_t$ explicitly but
it is suppressed in comparison to the leading term which
is proportional to $1/\G_t$ as the leading term in the 
denominator of  eq.~(\ref{phidef}) for the free quark 
Green function).
This divergent term is of the same nature as the divergence in the
$P$ wave amplitude discussed in Section~3.2. An accurate
calculation of this term can be done only within the
relativistic approximation. In contrast to the $P$ wave production
this term is parametrically suppressed
above the threshold
in comparison to the nonrelativistic continuum contribution
at least by the factor $\sqrt{\G_t/m_t}$ at $E\sim 0$
and can be safely 
omitted. However, it becomes important below
the resonance region when the nonrelativistic contribution
becomes small. Moreover, the denominator in
the right hand side of eq.~(\ref{phidef}) 
decreases rapidly below
the ground state pole. Therefore
a small uncertainty in the numerator 
would lead to a large uncertainty in 
the function $\Phi(k)$ and a reliable estimate of 
its numerical value is not possible 
in this region within the nonrelativistic 
approximation.             
Strictly speaking the accurate 
determination of the function $\Phi$ below the ground state pole 
requires the calculation of the relativistic $O(\G_t)$ 
contribution to the $S$ wave cross section (the denominator of 
eq.~(\ref{phidef})) which is not usually
considered since it does not 
lead to the divergence in the nonrelativistic expression.

\section{Discussion.}
The results of the numerical analysis 
for the physical observables
based on the
obtained analytical expressions are plotted 
in Figs.~1-4.

The constant $\tilde c^{++}_2$ appearing in the
hard  coefficient
$C^{++}(\al_s)$ in the $O(\al_s^2)$ order remains unknown.
The calculation of this parameter is necessary for the formal 
completion of the NNLO analysis. 
To find 
its numerical value one has to 
compute the $O(\al_s^2)$ perturbative 
QCD correction to the $\g\g$ cross section near the threshold in the 
formal limit $\al_s\ll \bt\ll 1$ and compare it with $O(\al_s^2)$ term 
in eq.~(\ref{rg}).  
In the case of $e^+e^-$ annihilation, however, the 
analogous contribution parameterized by $\tilde c_2^v$ is 
relatively small (about $10\%$ of the total NNLO correction) 
and the 
correction 
to the physical observables
in NNLO is saturated with the soft part 
of the total contribution 
determined by the 
corrections to the parameters of the nonrelativistic Green
function. Thus one can reasonably hope that the similar
situation can also take place 
for $\g\g$ collisions.
However the importance of this parameter for physical observables
is not crucial, it affects only the overall 
normalization of the cross sections. 
For example, it does not shift
the position of the resonance 
which is an important characteristic of the production 
and does not enter the ratio
$R^{++}(E)/R^{++}(0)$. For the numerical analysis of the cross 
section $R^\g$ we set $\tilde c_2^{++}=0$.

In our approach 
we deal with the soft corrections by summing them 
into the energy denominators of the discrete part of the Green 
function. In other words we treat the soft 
corrections as effective corrections to the parameters 
of the Green 
function written in a fixed functional form. 
The same 
approach has been advocated in refs.~\cite{BSS,Mel,Nag,HoaTeu,Yak} 
where all the 
corrections to the Green function have been found (numerically or 
analytically) in the 
form (\ref{endenom}). 
In ref.~\cite{Hoang}, however, a part of the NNLO 
corrections has not been resummed to the energy denominators of the 
discrete part of the Green function.  
On the other hand, in 
refs.~\cite{Mel,Nag,HoaTeu,Yak} 
the Schr{\"o}dinger equation~(\ref{Schr}) has been 
solved numerically, {\it i.e.} the NLO and NNLO correction to the Coulomb 
Hamiltonian have been taken into account
effectively in all orders of the nonrelativistic
series~(\ref{totcorr}) for the Green function
while we work strictly in NNLO. 
Our formulae reproduce the numerical 
result for $R^v$ of the most recent numerical analysis
\cite{Nag,HoaTeu,Yak} with $1\%-3\%$ accuracy that can be assigned
to the contribution of the higher  iterations of the 
NLO and NNLO corrections to the potential in
eq.~(\ref{totcorr}) beyond NNLO.

For the total cross sections 
which are dominated by $S$ wave contribution
we find the typical size of the
NNLO corrections to be of the order of $20\%$
in the overall normalization of the cross sections
and $\sim 40\%$ in the resonance energies expressed
in terms of the top quark pole mass, {\it i.e.}
of the order of the NLO ones (see Fig.~1).
Though 
the inclusion of the NLO corrections leads to 
a considerable
stabilization of the theoretical results for the cross sections
against changing the normalization point, 
the NNLO corrections
do not lead to better stability
as compared to NLO. In the overall
normalization of the cross sections the NLO and NNLO
corrections cancel each other to a large extent
while
the NLO and NNLO corrections to the resonance energies
are of the same sign and shift the resonance 
farther from the threshold.
They also make the peak more distinguishable which is the main 
difference between the leading 
Coulomb and NNLO approximations.

The leading order approximation for $R^e$
and $R^\g$ cross sections are the same
up to the normalization factor $2q_t^2$.
Up to the overall factor the difference between the cross sections
is determined by NNLO QCD and relativistic corrections
(see Fig.~2).
Above the threshold this difference
is determined by  
the difference between $B^{++}$ and $B^v$
coefficients and between $P$ wave contributions to
eqs.~(\ref{Ree},~\ref{Rgg})
{\it i.e.} by the pure relativistic corrections.
Below the threshold in the resonance region
this difference 
is determined also by
$A^i$ coefficients and is quite
sensitive to the value of $\al_s$.

Though using an infrared safe mass parameter instead 
of the pole mass   improves  the convergence of the series  
for the resonance energies it does not  affect the huge
NNLO corrections to the cross sections normalization. 
Moreover, it is not clear if there exist physically 
motivated mass and strong coupling parameters 
providing fast uniform convergence of the perturbative 
expansion for the cross sections in the threshold region. 
The absence of such a parameterization would mean the 
unavoidable significance of the high orders terms of
the threshold expansion. Some high order effects have been 
already considered in the literature.
The leading logarithmic corrections related to the 
renormalization group evolution of the hard coefficient
$C^{v}$ have been computed  \cite{BSS}.
The corresponding corrections to  the  $R^v$ cross section are $\pm 5\%$.     
In ref.~\cite{Nag} the running of the strong coupling constant
has been taken into account by introducing the energy dependent
soft normalization point of $\al_s$ entering the Coulomb potential   
in the numerical solution of the 
Schr{\"o}dinger equation. The resummation of the renormalization
group logarithms has an essential (up to $10\%$)
effect in the resonance region and reduces the normalization scale
dependence of the result. Furthermore, the  effect of retardation 
which introduces a new type of contributions absent in NLO and NNLO 
has been analyzed  for the low lying resonances \cite{KniPen}. 
The characteristic scale of the  leading ultrasoft contribution 
was found to be  about $-5 \%$ for the square of the ground state 
wave function at the origin and $+100~{\rm MeV}$ for the
ground state pole position.  

The result for the axial coupling contribution to
$e^+e^-\rightarrow t\bar t$ cross section is in a good agreement
with the numerical analysis of ref.~\cite{KuhTeu}. 
Up to the trivial normalization this contribution  coincides with the 
cross section $R^{+-}$ (Fig.~3).
Numerically it does not exceed $2\%$ of the total cross section 
and less than the
uncertainty due to the normalization scale dependence.  

The cross section $R^{+-}$ and the function $\Phi(k)$
obtain no contribution from 
the ground state resonance and, therefore 
they are
rather smooth because the top quark width 
smears the higher resonance contributions
very efficiently  
(Figs.~3,~4). These quantities
is rather insensitive to a variation of the normalization scale.
A typical NLO correction to $R^{+-}$ is about $10\%$
while the one to $\Phi(k)$ is about $15\%$ (the corrections to the
forward-backward asymmetry and top quark polarization
include also the hard normalization factors which 
have not been included to
$\Phi(k)$ itself and  the nonfactorizable corrections 
corrections discussed in Section~2.4).
Our result for the function
$\Phi(k)$ (Fig.~4) 
is in a good agreement
with the results of numerical analysis \cite{Sum,Har} for the
energies above the ground state resonance. 
There is some discrepancy
between the results below the resonance. 
However the reliable
estimate of the function $\Phi$ is not possible 
in this region with the pure nonrelativistic treatment
of the top quark width as has been explained in
Section~3.3.

The final remark of this section concerns the 
optimal choice of the  normalization and factorization scales.
The hard scale appears in the hard coefficients as $\ln{(m_t/\mu_h)}$
{\it i.e.} the typical hard scale of the problem is the top
quark mass.  Though in a fixed order of the perturbative
expansion the hard coefficients do not depend on $\mu_h$ 
one can put  $\mu_h \sim m_t$ to minimize the potentially
large logarithmic contributions of the higher order terms.
In practice the NNLO results are almost independent 
of $\mu_h$ when $\mu_h \sim m_t$.
On the other hand the requirement of convergence of the perturbative expansion
around the Coulomb Green function
restricts the allowed range for the choice of a
soft normalization point which can be used for reliable estimates.
The soft physical scale of the problem
is determined by the natural infrared cutoff related to the top quark width
$\sqrt{m_t\G_t}$ that measures the distance to the nearest
singularity in the complex energy plane and/or by the characteristic
scale of the Coulomb problem $\lambda$  {\it i.e.}
$\mu_s\sim 15~{\rm  GeV}$. Both scales are rather close to each
other for the case of top quark that makes
possible a uniform description of both perturbative QCD and Coulomb
resonance effects. Indeed, for $\mu_s\sim 15~{\rm GeV}$
the soft NLO correction, for example, to the energy level~(\ref{e2})
reaches its minimal magnitude. However, at this scale the
NNLO correction exceeds the NLO one  and the series for the energy levels
seems to diverge. Moreover, for such a low soft normalization 
point the  NNLO corrections to the wave function at the origin
which cannot be dumped by the quark mass redefinition become uncontrollable.  
This is not surprising since the normalization
scale is defined in a rather
artificial $\overline {\rm MS}$ scheme that has little to do with
peculiarities of $t\bar t$ physics and
there is no reason for a literal coincidence of $\mu_s$
parameter with any physical scale of the process.
The relative weight of the NNLO correction to the
Green function as well as
the dependence of the cross sections on $\mu_s$ is stabilized at
$\mu_s$\raisebox{-3pt}{$\stackrel{>}{\sim}$}$40~{\rm GeV}$
which can be considered as an optimal choice of the soft
normalization point. The price one pays for 
using  different soft and hard normalization scales
is the incomplete cancelation of the factorization scale 
dependence but this effect is suppressed 
by an additional power of $\al_s$. 
Another source of the  dependence on the  factorization scale is  the 
factorized form~(\ref{Rv},~\ref{Rpp}) of the cross sections where some 
higher order $\mu_f$-dependent terms are kept. The numerical analysis, 
however, shows that  the results  are  rather insensitive 
to the factorization scale chosen in the region  $\mu_f\sim m_t$.

\section{Conclusion}
The basic observables of the top quark pair production
in $e^+e^-$ annihilation and $\gamma\gamma$ collision
have been considered in the threshold region.
The threshold effects are described by three universal
functions related to the $S$, $P$ wave production and
$S-P$ wave interference which have been computed analytically
within (potential) NRQCD.
An explicit analytical expression for the soft part of
the NNLO corrections to the total cross section has been 
obtained.
The $e^+e^-\rightarrow t\bar t$ threshold cross section
has been obtained in NNLO in 
the closed form including the contribution
due to the top quark axial coupling. The forward-backward
asymmetry of the quark-antiquark pair
production in $e^+e^-$ annihilation  and top quark polarization
in both processes have been computed analytically up to NLO.
The running of the strong
coupling constant and the finite top quark width effects 
in the $P$ wave
production and $S-P$ wave interference have been
taken into account
properly  within the analytical approach.

In combination, these uncorrelated observables
form an efficient tool for investigating quark interactions.
As  independent sources they can also be used for
determination of the
theoretical uncertainty in the numerical values of the strong
coupling constant $\al_s$,
the top quark mass, and the top quark width
extracted from the experimental date on top-antitop production.

The high order corrections turn out to be relatively large for all
observables and important for 
the accurate description of the top quark
physics near production threshold.

\vspace{5mm}
\noindent
{\large \bf Acknowledgments}\\[2mm]
This work is partially supported
by Volkswagen Foundation under contract
No.~I/73611 and Russian Fund for Basic Research under contract
No.~97-02-17065. The work of A.A.Penin is supported in part  by
the  Russian Academy of Sciences Grant No.~37. The present stay
of A.A.Pivovarov in Mainz, where the paper has been
completed, was made possible due to Alexander von Humboldt fellowship.
We are thankful to K.~Melnikov and A.~Czarnecki  \cite{CzaMel}
for pointing out an error in the partial wave decomposition
of $R^{++}$ in the previous version of this paper.

\section*{Appendix.}
{\large\bf A.} The correction $\Delta_2^{(2)}G$ due to the $\Delta_2V$
part of the potential has the form
\cite{KPP}
\[ \Delta_2^{(2)}G=\left({\al_s\over
4\pi}\right)^2{C_F\al_sm_t^2\over 4\pi}\left( \sum_{m=0}^\infty F(m)^2
\left((m+1)\left(C_0^2+L(k) C_1^2+L(k)^2 C_2^2\right)
\right.\right.
\]
$$
\left.
+(m+1)\Psi_1(m+2)\left(C_1^2+2L(k)
C_2^2\right)+K(m)C_2^2\right)
$$
\[
+2\sum_{m=1}^\infty\sum_{n=0}^{m-1}
F(m)F(n)\left(-
{n+1\over m-n}\left(C_1^2 +2L(k) C_2^2\right)
+K(m,n)C_2^2\right)
\]
\[
+2\sum_{m=0}^\infty F(m)
\left(C_0^2+L(k) C_1^2+
(L(k)^2+J(m))C_2^2-(2\gamma_E+
\Psi_1(m+1))
\left(C_1^2+2L(k) C_2^2\right)\right)
\]
$$
\left.+L(k)C_0^2+\left(-\gamma_E L(k)+
{1\over 2}L(k)^2\right)
C_1^2
+I(k)C_2^2\right)
$$
where
\[
L(k)=\ln\left({\mu_s\over 2k }\right)
\]
\[
K(m)
=
(m+1)\left(\Psi_1(m+2)^2-\Psi_2(m+2)+{\pi^2\over3}-{2\over(m+1)^2}\right)
\]
$$
-2(\Psi_1(m+1)+\gamma_E),
$$
$$
K(m,n)= 2{n+1\over m-n}\left(\Psi_1(m-n)-{1\over n+1}+2\gamma_E\right)
$$
\[
+2{m+1\over m-n}(\Psi_1(m-n+1)-\Psi_1(m+1)),
\]
$$
J(m)=2(\Psi_1(m+1)+\gamma_E)^2+\Psi_2(m+1)-\Psi_1(m+1)^2+2\gamma_E^2,
$$
$$
I(k)=\left(\gamma_E+{\pi^2\over 6}\right)L(k)
-\gamma_E L(k)^2+{1\over 3}
L(k)^3.
$$
The correction $\Delta_2^{(1)}G$ due to the
second iteration of $\Delta_1V$ term \cite{PP}
\[
\Delta_2^{(1)}G=\left({\al_s\over 4\pi}\right)^2
{(C_F\al_s)^2\over 4\pi}{m_t^3\over 2k}
\left( \sum_{m=0}^\infty H(m)^3(m+1)
\right.
\]
$$
\left(C_0^1+
\left(\Psi(m+2)+
L(k)\right)C_1^1\right)^2
$$
\[
-2\sum_{m=1}^\infty\sum_{n=0}^{m-1}{n+1\over m-n}C_1^1
\left(H(m)^2H(n)\left(C_0^1+\left(\Psi(m+2)+
L(k)-{1\over 2}{1\over m-n}\right)C_1^1\right)
\right.
\]
\[
\left.
+H(m)H(n)^2\left(C_0^1+\left(\Psi(n+2)+
L(k)-{1\over 2}{n+1\over (m-n)(m+1)}\right)C_1^1\right)\right)
\]
\[
+2(C_1^1)^2\left(\sum_{m=2}^\infty\sum_{l=1}^{m-1}\sum_{n=0}^{l-1}
{H(m)H(n)H(l)}{n+1\over (l-n) (m-n)}\right.
\]
$$
+\sum_{m=2}^\infty\sum_{n=1}^{m-1}\sum_{l=0}^{n-1}
{H(m)H(n)H(l)}{l+1\over (n-l)(m-n)}
$$
\[
\left.\left.
+\sum_{n=2}^\infty\sum_{m=1}^{n-1}\sum_{l=0}^{m-1}
{H(m)H(n)H(l)}{(l+1)(m+1)\over (n+1)(n-l)(n-m)}\right)\right)
\]
where
\[
H(m)={1\over m+1-\nu}
\]

\vspace{5mm}
\noindent
{\large\bf B.} 
We define dimensionally regularized value of 
the Coulomb Green function at the origin
directly through the relation
\[
G^{d.r.}_C(0,0,k)=\int d^dp \tilde G(p,k)
\]
with $d=3-2\ve$. Using the following
representation of the momentum space Green function
\[
\tilde G(p,k)={m_t\over 8\pi^3} \int_0^\infty \left({1+t\over t}\right)^\nu dt
{4 k^2 (1+2t)\over (p^2 + k^2(1+2t)^2)^2}
\]
one obtains
\[
G^{d.r.}_C(0,0,k)={m_t k\over 2\pi}\left(\frac{\mu_f}{k}\right)^{2\ve}
\int_0^\infty \left({1+t \over t}\right)^\nu
{dt\over (1+2t)^{2\ve}}
\]
where we omit inessential factors related to the precise definition 
of integration measure in $d$ dimensions. These factors lead to  
the multiplication of the Green function with an additional 
quantity $1+O(\ve)$
and  
can be taken into account by the redefinition 
of $\mu_f$ scale.
The integral in the right hand side of this equation is
\[
\int_0^\infty \left({1+t\over t}\right)^\nu
{dt\over (1+2t)^{2\ve}}=2^{-2\ve}
B(-1+2\ve,1-\nu){}_2F_1(2\ve,-1+2\ve;2\ve-\nu;{1\over 2})
\]
where $B(z,w)$ is the Euler $B$-function and $_2F_1(a,b;c;z)$
is the hypergeometric function. 
Upon expanding the above expression in
$\ve$ around $\ve=0$ one arrives
at the final result for the dimensionally regularized Coulomb Green
function.
The factorization scale 
$\mu_f$ in eq.~(\ref{dimreg}) is chosen in such a way that it is true
as written.
Note that the Green function regularized in this way 
does not automatically 
match the hard coefficient computed in ${\overline{\rm MS}}$
scheme of the orthodox dimensional regularization \cite{c1,c2}. 

\vspace{5mm}
\noindent
{\large\bf C.} The NNLO corrections to the square
of the  Coulomb   $^3S_1$ and  $^1S_0$ heavy quark
bound state wave function at the origin have the form \cite{PP1}
\[
\Delta_{k^2}\psi^2_{0m}=B^i{C_F^2\al_s^2\over 4(m+1)^2},
\]
\[
\Delta_{{\bf \Delta}^2,NA,BF}\psi^2_{0m}=-{C_F^2\al_s^2}
\left(
{15\over 8}{1\over(m+1)^2}+\left({5-A^i\over 2}+{C_A\over C_F}\right)
\left(-\ln\left({\mu_{f}(m+1)\over \lambda}\right)\right.\right.
\]
\[
\left.\left.
+\gamma_E
+\Psi_1(m+1)-{1\over(m+1)}\right)
\right),
\]
\[
\Delta^{(2)}_2\psi^2_{0m}=\left({\alpha_s\over 4\pi}\right)^2
\left(3(C^2_0+\bar L(m)C_1^2+\bar L(m)^2C^2_2) +
(-1-2\gamma_E+{2\over m+1}+\Psi_1(m+2)\right.
\]
\[
-2(m+1)\Psi_2(m+1))(C_1^2+2\bar L(m)C^2_2)+\left({K(m)\over m+1}+
2K(m)-2\Psi_1(m+2)\right.
\]
\[
\left.\left.
+2\sum_{n=0}^{m-1}{m+1\over (n-m)(n+1)}K(m,n)+2\sum_{n=m+1}^\infty
{m+1\over (n-m)(n+1)}K(n,m)
\right)C_2^2
\right),
\]
\[
\Delta^{(1)}_2\psi^2_{0m}=\left({\alpha_s\over 4\pi}\right)^2
\left(3(C_0^1+(\bar L(m)+
\Psi_1(m+2))C_1^1)^2\right.
\]
\[
+2C_1^1\left(\sum_{n=0}^{m-1}{(n+1)(m+1)\over (n-m)^3}\left(
C_0^1+\left((\bar L(m)+
\Psi_1(n+2)+{1\over 2}{n+1\over (n-m)(m+1)}\right)C_1^1\right)
\right.
\]
\[
\left.
-\sum_{n=m+1}^{\infty}{(m+1)^2\over (n-m)^3}\left(
C_0^1+\left((\bar L(m)+
\Psi_1(n+2)-{1\over 2}{1\over n-m}\right)C_1^1\right)
\right)
\]
\[
+\left.
2C_1^1\left(C_0^1+\left(\bar L(m)+\Psi_1(m+2)\right)
C_1^1\right)
\left(-{5\over 2}+\sum_{n=0}^{m-1}{n+1\over (n-m)^2}U(m,n)
\right.
\right.
\]
\[
\left.
-\sum_{n=m+1}^{\infty}{m+1\over (n-m)^2}U(m,n)
\right)
+2(C_1^1)^2\left({1\over 2}-\sum_{n=0}^{m-1}{n+1\over (n-m)^2}
+\sum_{n=m+1}^{\infty}{m+1\over (n-m)^2}\right.
\]
\[
\left.
+{1\over 2}\sum_{n=0}^{m-1}{n+1\over (n-m)^3}U(m,n)
+{1\over 2}\sum_{n=m+1}^{\infty}{(m+1)^2\over (n-m)^3(n+1)}U(m,n)
\right.
\]
\[
+\sum_{n=1}^{m-1}\sum_{l=0}^{n-1}\left(
{(l+1)(n+1)\over (n-m)^2(l-m)^2}-{(l+1)(m+1)\over (n-m)^2(l-m)(n-l)}-
{(l+1)(m+1)\over (l-m)^2(n-m)(n-l)}\right)
\]
\[
+\sum_{n=m+1}^{\infty}\sum_{l=0}^{m-1}\left(
-{(l+1)(m+1)\over (n-m)^2(l-m)^2}+{(l+1)(m+1)^2\over (n-m)^2(l-m)(n-l)(n+1)}
\right.
\]
\[
\left.
-{(l+1)(m+1)\over (l-m)^2(n-m)(n-l)}\right)
+\sum_{n=2}^{\infty}\sum_{l=m+1}^{n-1}\left(
{(m+1)^2\over (n-m)^2(l-m)^2}\right.
\]
\[
\left.\left.\left.
+{(l+1)(m+1)^2\over (n-m)^2(l-m)(n-l)(n+1)}+
{(m+1)^2\over (l-m)^2(n-m)(n-l)}\right)\right)\right)
\]
where $\bar L(m)=L(\lambda /(m+1))$ and
\[
U(m,n)=3+{n+1\over m+n+2}
-2{(n+1)^2\over (n-m)(n+m+2)}
\]

\vspace{5mm}
\noindent
{\large\bf D.} The NNLO corrections to the   Coulomb  $^3S_1$ and  $^1S_0$
heavy quark bound state energy levels \cite{MY,PP1,PY}
\[
\Delta_{{\bf \Delta}^2,NA,BF}E_{0m}={C_F^2\al_s^2\over (m+1)}
\left({C_A\over C_F}+{5-A^i\over 2} -{11\over 16}{1\over (m+1)}\right),
\]
\[
\Delta_2^{(2)}E_{0m}=2\left({\alpha_s\over 4\pi}\right)^2
\left(C_0^2+\bar L(m)C_1^2+\bar L(m)^2C_2^2+
\Psi_1(m+2)(C_1^2+2\bar L(m)C_2^2)\right.
\]
\[
\left.
+{K(m)\over (m+1)}C_2^2\right),
\]
\[
\Delta_2^{(1)}E_{0m}=\left({\alpha_s\over 4\pi}\right)^2
\left(\left(C_0^1+(\bar L(m)-2+
\Psi_1(m+2))C_1^1\right)
\left(C_0^1+(\bar L(m)
+\Psi_1(m+2))C_1^1\right)\right.
\]
\[
\left.
+\left({2\over  (m+1)}(\gamma_E+\Psi_1(m+2))
-2\Psi_2(m+1)-(m+1)\Psi_3(m+1)\right)(C_1^1)^2\right),
\]

\vspace{5mm}
 
\section*{Figure captions}

\noindent
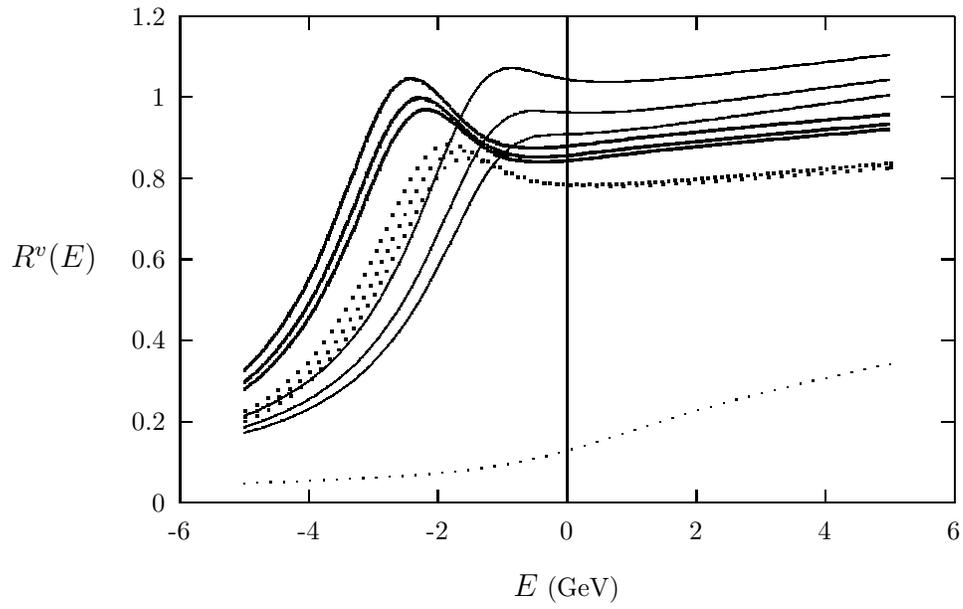
{\bf Fig. 1.}
The normalized  cross section  $R^v(E)$
in the leading order (solid lines),
NLO (bold dotted lines) and
NNLO (bold solid lines) for $m_t=175~{\rm GeV}$, $\G_t=1.43~{\rm GeV}$,
$\al_s(M_Z)=0.118$ and $\mu_s=50~{\rm GeV},~75~{\rm GeV}$
and $100~{\rm GeV}$.
The dotted line corresponds to the result in Born approximation.

\noindent
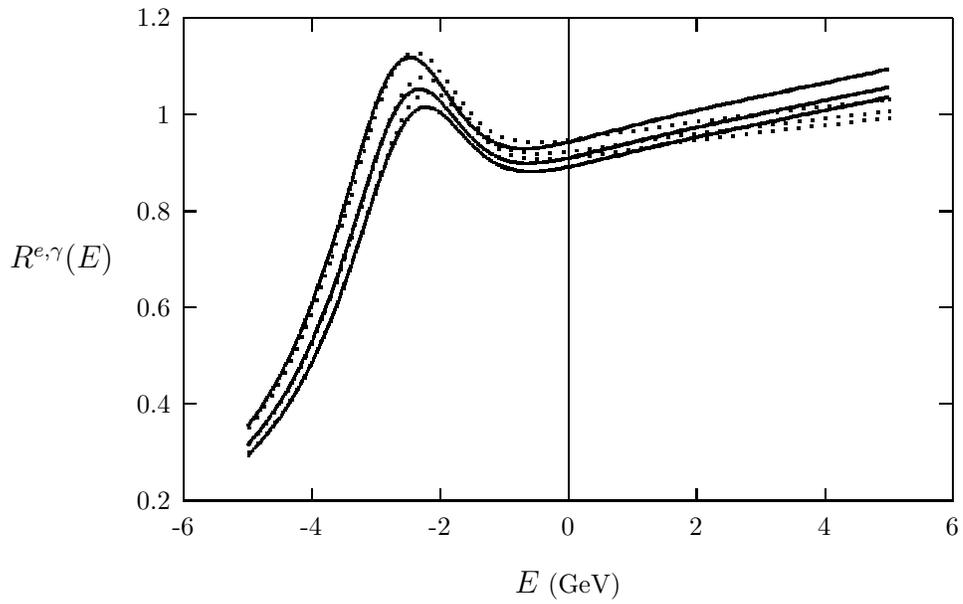
{\bf Fig. 2.}
The normalized cross sections $R^e(E)$ (dotted lines) and $R^\g(E)$
(solid lines) in NNLO for $\tilde c_2^{++}=0$, $m_t=175~{\rm GeV}$,
$\G_t=1.43~{\rm GeV}$,
$\al_s(M_Z)=0.118$, $\sin^2\theta_W=0.232$, $M_Z=91.2~{\rm GeV}$
and $\mu_s=50~{\rm GeV},~75~{\rm GeV}$ and $100~{\rm GeV}$.

\noindent
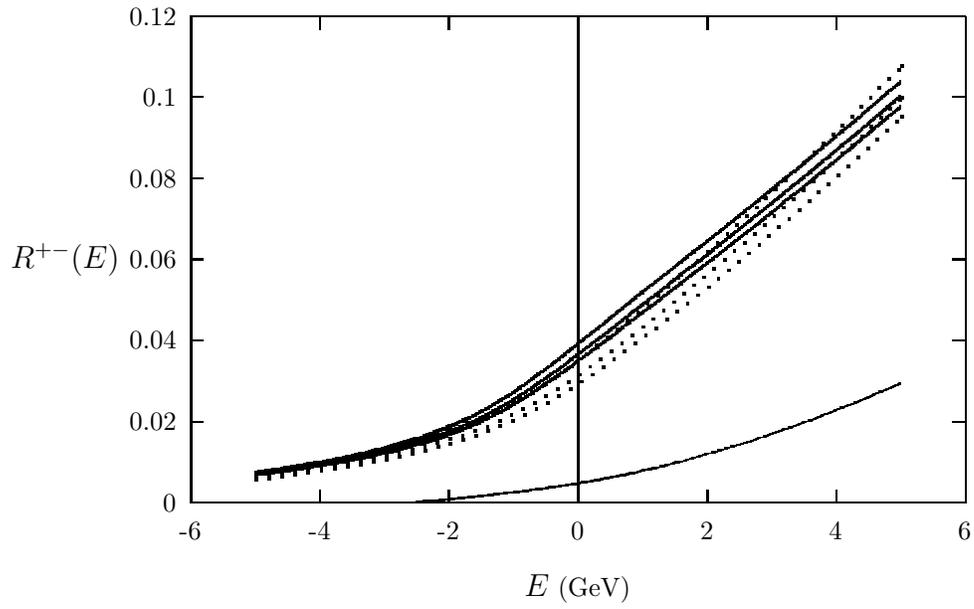
{\bf Fig. 3.}
The normalized cross section  $R^{+-}(E)$
in the leading order (dotted lines)  and
NLO (bold solid lines) for $m_t=175~{\rm GeV}$, $\G_t=1.43~{\rm GeV}$,
$\al_s(M_Z)=0.118$ and $\mu_s=50~{\rm GeV},~75~{\rm GeV}$
and $100~{\rm GeV}$.
The solid line corresponds to the result in Born approximation.

\noindent
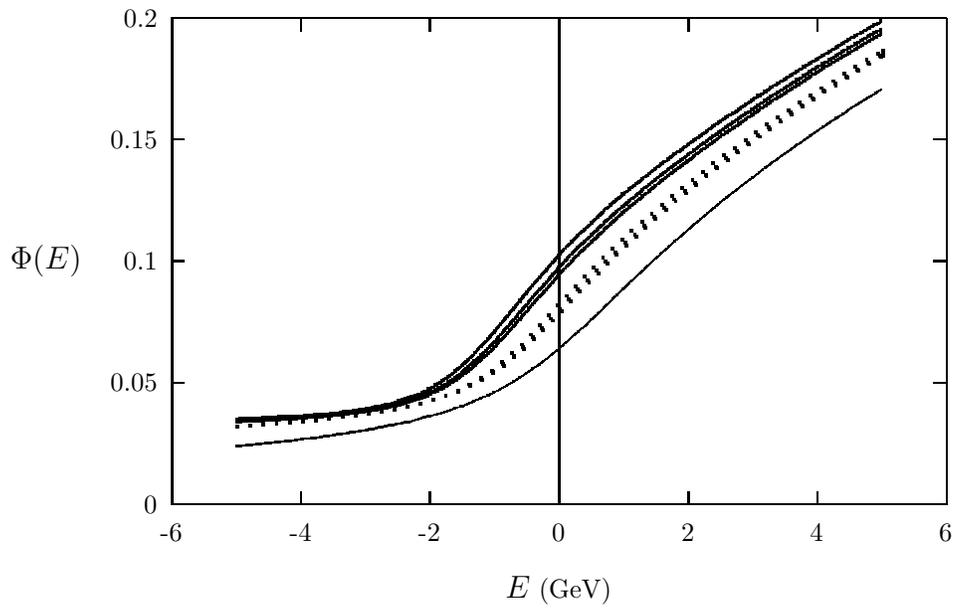
{\bf Fig. 4.}
The function  $\Phi(E)$
in the leading order (dotted lines)  and
NLO (bold solid lines) for $m_t=175~{\rm GeV}$, $\G_t=1.43~{\rm GeV}$,
$\al_s(M_Z)=0.118$ and $\mu_s=50~{\rm GeV},~75~{\rm GeV}$
and $100~{\rm GeV}$.
The solid line corresponds to the result in Born approximation.

\newpage

\vspace{10mm}
\begin{center}

\setlength{\unitlength}{0.240900pt}
\ifx\plotpoint\undefined\newsavebox{\plotpoint}\fi


\vspace{5mm}
{\bf Fig. 4.}
\end{center}


\begin{thebibliography}{99}

\bibitem{FK}   V.S.Fadin and V.A.Khoze, Pis'ma Zh.Eksp.Teor.Fiz.
               {\bf 46}(1987)417; \\ Yad.Fiz. {\bf 48}(1988)487.

\bibitem{ttgg} V.S.Fadin and V.A.Khoze, Yad.Fiz. {\bf 93}(1991)1118;\\
               I.I.Bigi, V.S.Fadin  and V.A.Khoze, Nucl.Phys. {\bf
               B377}(1992)461;\\
               I.I.Bigi, F.Gabbiani and V.A.Khoze, Nucl.Phys.
               {\bf B406}(1993)3;\\
               J.H.K\"uhn, E. Mirkes and J.Steegborn, Z.Phys. 
               {\bf C57}(1993)615. 

\bibitem{ttee} W.Kwong, Phys.Rev. {\bf D43}(1991)1488;\\
               M.J.Strassler and M.E.Peskin,  Phys.Rev.
               {\bf D43}(1991)1500;\\
               M.Jezabek,  J.H.K{\"u}hn and T.Teubner, Z.Phys.
               {\bf C56}(1992)653;\\
               Y.Sumino {\it et al.},  Phys.Rev. {\bf D47}(1993)56.

\bibitem{exp}  E.Accomando {\it et al.}, Phys.Rep. {\bf 299}(1998)1.

\bibitem{Sum} H.Murayama and  Y.Sumino   Phys.Rev. {\bf 47}(1993)82;

\bibitem{Har} R.Harlander, M.Jezabek,  J.H.K{\"u}hn and T.Teubner,
               Phys.Lett. {\bf B346}(1995)137.

\bibitem{FKK} V.S.Fadin, V.A.Khoze and M.I.Kotsky, Z.Phys {\bf C64}(1994)45.


\bibitem{CasLep} W.E.Caswell and G.E.Lepage, Phys.Lett.
                 {\bf B167}(1986)437;\\
                 G.E.Lepage {\it et al.},  Phys.Rev. {\bf D46}(1992)4052;\\
                 G.T.Bodwin, E.Braaten and G.P.Lepage,
                 Phys.Rev. {\bf D51}(1995)1125.           

\bibitem{Man}  A.V.Manohar,  Phys.Rev. {\bf D56}(1997)230.

\bibitem{LukMan} M.Luke and  A.V.Manohar,  Phys.Rev. {\bf D55}(1997)4129.

\bibitem{GriRot} B.Grinstein and I.Z.Rothstein,  Phys.Rev. {\bf D57}(1998)78.

\bibitem{LukSav} M.Luke and  M.J.Savage,  Phys.Rev. {\bf D57}(1998)413.

\bibitem{PinSot} A.Pineda and J.Soto,
                  Nucl.Phys.Proc.Suppl. {\bf 64}(1998)428.
   
\bibitem{BenSmi} M. Beneke and V.A. Smirnov,
                 Nucl.Phys.  {\bf B522}(1998)321.

\bibitem{Lab} P.Labelle, Phys.Rev. {\bf D57}(1998)093013.

\bibitem{Gr}   H.W.Griesshammer, Preprint {\bf UW-98-22}, hep-ph/9810235.

\bibitem{H1}    A.H.Hoang, Phys.Rev. {\bf D56}(1997)5851.

\bibitem{KPP}  J.H.K{\"u}hn, A.A.Penin and A.A.Pivovarov,
               Nucl.Phys. {\bf B534}(1998)356.

\bibitem{PP} A.A.Penin and A.A.Pivovarov, Phys.Lett. {\bf B435}(1998)413.

\bibitem{MY} K.Melnikov and A.Yelkhovsky, Phys.Rev. {\bf D59}(1999)114009.

\bibitem{PP1} A.A.Penin and A.A.Pivovarov, Nucl.Phys. {\bf B549}(1999)217.

\bibitem{BSS} M.Beneke, A.Singer and V.A.Smirnov,
              Phys.Lett. {\bf B454}(1999)137.      

\bibitem{BenSin} M.Beneke and A.Singer,
                 Preprint {\bf CERN-TH/99-163}, hep-ph/9906475.    

\bibitem{H2}    A.H.Hoang, Preprint {\bf CERN/TH 99-152}, hep-ph/9905550. 

\bibitem{KniPen} B.A.Kniehl and   A.A.Penin,
                 Preprint {\bf DESY 99-099},  hep-ph/9907489.

\bibitem{Hoang} A.H.Hoang and T.Teubner,
                Phys.Rev. {\bf D58}(1998)114023.

\bibitem{Mel} K.Melnikov and A.Yelkhovsky, Nucl.Phys. {\bf B528}(1998)59.

\bibitem{KuhTeu} J.H.K\"uhn and T.Teubner,  Eur.Phys.J. {\bf C9}(1999)221.

\bibitem{Nag}  T.Nagano, A.Ota and  Y.Sumino, 
               Phys.Rev. {\bf D60}(1999)114014.

\bibitem{HoaTeu} A.H.Hoang and T.Teubner,   
                  Phys.Rev. {\bf D60}(1999)114027.
                 
\bibitem{PP2}   A.A.Penin and A.A.Pivovarov, Nucl.Phys. {\bf B550}(1999)375.

\bibitem{c1}  A.Czarnecky and K.Melnikov,
              Phys.Rev.Lett. {\bf 80}(1998)2531.

\bibitem{c2}  M.Beneke, A.Signer and V.A.Smirnov,
              Phys.Rev.Lett. {\bf 80}(1998)2535.

\bibitem{PinSot1} A. Pineda and J. Soto, Phys.Lett. {\bf B420}(1998)391,
                  Phys.Rev. {\bf D59}(1999)016005.

\bibitem{CMY} A. Czarnecki, K. Melnikov and A. Yelkhovsky,
              Phys.Rev. {\bf A59}(1999)4316.

\bibitem{Kuh}  J.H.K\"uhn, {\it ``Theory of the top quark production
               and decay''}, Lectures given at XXXIII SLAC Summer Institute
               on Particle Physics {\it ``The Top Quark and the Electroweak
               Interaction''}

\bibitem{Kar}  R.Karplus and A.Klein  Phys.Rev. {\bf 87}(1952)848;\\
                G.K{\"a}llen and A.Sarby,
                K.Dan.Vidensk.Selsk.Mat.-Fis.Medd. {\bf 29}(1955), N17, 1;\\
                R.Barbieri {\it et al.}, Phys.Lett. {\bf B57}(1975)535.

\bibitem{RRY}  L.J.Reinders, H.R.Rubinstein and S.Yazaki,
               Nucl.Phys. {\bf B181}(1981)109;\\
               J.J{\'e}rzak, E.Laemann and  P.M.Zerwas,
               Phys.Rev. {\bf D25}(1982)1218.

\bibitem{HarBr}  I.Harris and L.M.Brown, Phys.Rev. {\bf 105}(1957)1656;\\
                 R.Barbieri {\it et al.}, Nucl.Phys. {\bf B154}(1979)535.

\bibitem{BarKuh} R.Barbieri {\it et al.}, Nucl.Phys. {\bf B192}(1981)61;\\
                 J.H.K\"uhn and P.M.Zerwas, Phys.Rep. {\bf B167}(1988)321.

\bibitem{Piv}  A.A. Pivovarov, Phys.Rev. {\bf D47}(1993)5183.

\bibitem{Holl} B.Grzadkowski {\it et al. },  Nucl.Phys. {\bf B281}(1987)18;\\
               R.J.Guth and J.H.K\"uhn, Nucl.Phys. {\bf B368}(1992)38;\\
               W.Beenakker, W.Hollik and Van der Mark,
               Nucl.Phys. {\bf B365}(1991)24.

\bibitem{Denn} A.Denner, S.Dittmaier and M.Strobel,
               Phys.Rev. {\bf D53}(1996)44.

\bibitem{Chi} B.M.Chibisov and M.V.Voloshin, Mod.Phys.Lett.
              {\bf A13}(1998)973.

\bibitem{FKM}  V.S.Fadin, V.A.Khoze and A.D.Martin,
               Phys.Rev. {\bf D49}(1994)2247;\\
               K.Melnikov  and O.Yakovlev, Phys.Lett. {\bf B324}(1994)217.

\bibitem{PetSum} K. Fujii, T.Matsui and  Y.Sumino,  Phys.Rev. {\bf D50}(1994)4341;\\
                 R.Harlander, M.Jezabek, J.H.K\"uhn and
                 M.Peter, Z.Phys. {\bf C73}(1997)477;\\ 
                 M.Peter and Y.Sumino, Phys.Rev. {\bf D57}(1998)6912. 

\bibitem{Gup}  S.N.Gupta and S.F.Radford,  Phys.Rev. {\bf D24}(1981)2309;
               Phys.Rev. \\{\bf D25}(1982)3430 (Erratum);\\
               S.N.Gupta, S.F.Radford and W.W.Repko,
               Phys.Rev. {\bf D26}(1982)3305.

\bibitem{Landau} L.D.Landau and E.M.Lifshitz, Relativistic Quantum
                 Theory, Part 1 (Pergamon, Oxford, 1974).

\bibitem{Fish}   W.Fisher, Nucl.Phys. {\bf B129}(1977)157;\\
                 A.Billoire, Phys.Lett. {\bf B92}(1980)343.

\bibitem{Peter} M.Peter,  Phys.Rev.Lett. {\bf 78}(1997)602;
                Nucl.Phys {\bf B501}(1997)471;\\
                Y.Schr{\"o}der, Phys.Lett. {\bf B447}(1999)321.

\bibitem{sd}  A.H.Hoang, Phys.Rev. {\bf D56}(1997)7276.

\bibitem{INOK} K.A.Ispiryan {\it et al.}, Yad.Fiz. {\bf 11}(1970)1278.

\bibitem{PY} A.Pineda and F.J.Yndurain, Phys.Rev. {\bf D58}(1998)094022.

\bibitem{BB} M.Beneke and V.M.Braun Nucl.Phys. {\bf B426}(1994)301;\\ 
             I.I.Bigi {\it et al.},  Phys.Rev. {\bf D50}(1994)2234.
           
\bibitem{VL} M.B.Voloshin, Nucl.Phys. {\bf B154}(1979)365;\\
             H.Leutwyler, Phys.Lett. {\bf B98}(1981)447.

\bibitem{Yak} O.Yakovlev, Phys.Lett. {\bf B457}(1999)170.

\bibitem{CzaMel}  A.Czarnecki and K.Melnikov, Preprint 
{\bf SLAC-PUB-8972}, hep-ph/0108233. 




\end{thebibliography}
\end{document}